\documentclass[iop]{emulateapj}
\usepackage{epsf}
\usepackage{graphicx, commath, color, hyperref}
\def\dir_plots{./}

	\newcommand{\photoz}{photo-$z$}
	\newcommand{\photozs}{photo-$z$s}
	\newcommand{\specz}{spec-$z$}
	\newcommand{\speczs}{spec-$z$s}

	\newcommand{\Msun}{$M_{\odot}$}
	\newcommand{\msun}{M_{\odot}}

	\newcommand{\synmag}{{\sc synmag}}
  	\newcommand{\birth}{$b_{1000}$}

	\newcommand{\Mmgc}{\ensuremath{M_{*{\rm MGC}}}}  % M_MGC
	\newcommand{\MmgcOpt}{\ensuremath{M_{*{\rm MGC}}^{\rm opt}}}  % M_MGC Optical
	\newcommand{\Mised}{\ensuremath{M_{*{\rm iSED}}}}  % M_iSED
	\newcommand{\MisedFSPS}{\ensuremath{M_{*{\rm iSED}}^{\rm FSPS}}}  % M_iSED FSPS
	\newcommand{\MisedBC}{\ensuremath{M_{*{\rm iSED}}^{\rm BC03}}}  % M_iSED BC03
	\newcommand{\MisedMa}{\ensuremath{M_{*{\rm iSED}}^{\rm Ma05}}}  % M_iSED Maraston
	\newcommand{\Mvagc}{\ensuremath{M_{*{\rm VAGC}}}}  % M_iSED Maraston

\newcommand{\zbest}{$z_{\rm best}$}

\newcommand{\redmapper}{{\it redMaPPer}}  
\newcommand{\redmagic}{{\it redMaGiC}}

	\newcommand{\area}{$139.4$}   % area in degrees 
	\newcommand{\vol}{$0.3$}   % volume in Gpc^3 
	\newcommand{\cat}{{\sc s82-mgc}}   % number of BOSS spec-zs in final sample
	\newcommand{\kband}{reddest band}  % replacement for K-band
	\newcommand{\ukwide}{\textsc{ukwide}}  % replacement for K-band

\newcommand{\catpaper}{Paper I}   % Reference to Catalog Paper
\newcommand{\comppaper}{Paper II}   % Ref to completeness paper
\newcommand{\mfnpaper}{Paper III}   % Ref to mass function paper
\newcommand{\isedfit}{\texttt{iSEDfit}}

\shorttitle{S82-MGC III: A Lack of Growth Among Massive Galaxies}
\shortauthors{Bundy et al.}

\submitted{Accepted in ApJ}

\begin{document}

\title{The Stripe 82 Massive Galaxy Project III: A Lack of Growth Among Massive Galaxies}

\author{Kevin Bundy\altaffilmark{1,2}, Alexie Leauthaud\altaffilmark{2}, Shun Saito\altaffilmark{3}, Claudia Maraston\altaffilmark{4}, David A.~Wake\altaffilmark{5}, Daniel Thomas\altaffilmark{4}}

\altaffiltext{1}{UC Observatories, MS: UCO Lick, UC Santa Cruz, 1156 High St, Santa Cruz, CA 95064, USA}
\altaffiltext{2}{Department of Astronomy \& Astrophysics, UC Santa Cruz, 1156 High St, Santa Cruz, CA 95064, USA}
\altaffiltext{3}{Max-Planck-Institut f\"ur Astrophysik, Karl-Schwarzschild-Str. 1, 85741 Garching, Germany}
\altaffiltext{4}{Institute of Cosmology and Gravitation, University of Portsmouth, Portsmouth, UK}
\altaffiltext{5}{Department of Physical Sciences, The Open University, Milton Keynes, MK7 6AA, UK}

\begin{abstract}

The average stellar mass ($M_*$) of high-mass galaxies ($\log M_*/\msun > 11.5$) is expected to grow by $\sim$30\%
since $z \sim 1$, largely through ongoing mergers that are also invoked to explain the observed increase in galaxy
sizes.  Direct evidence for the corresponding growth in stellar mass has been elusive, however, in part because the
volumes sampled by previous redshift surveys have been too small to yield reliable statistics.  In this work, we make
use of the Stripe 82 Massive Galaxy Catalog (\cat{}) to build a mass-limited sample of 41,770 galaxies ($\log M_*/\msun
> 11.2$) with optical to near-IR photometry and a large fraction ($>$55\%) of spectroscopic redshifts.  Our sample
spans 139 deg$^2$, significantly larger than previous efforts. After accounting for a number of potential systematic
errors, including the effects of $M_*$ scatter, we measure galaxy stellar mass functions over $0.3 < z < 0.65$ and
detect no growth in the typical $M_*$ of massive galaxies with an uncertainty of 9\%.  This confidence level is
dominated by uncertainties in the star formation history assumed for $M_*$ estimates, although our inability to
characterize low surface-brightness outskirts may be the most important limitation of our study. Even among these
high-mass galaxies, we find evidence for differential evolution when splitting the sample by recent star formation (SF)
activity.  While low-SF systems appear to become completely passive, we find a mostly sub-dominant population of
galaxies with residual, but low rates of star formation ($\sim$1 \Msun/yr) number density does not evolve.
Interestingly, these galaxies become more prominent at higher M$_*$, representing $\sim$10\% of all galaxies at
$10^{12} \msun$ and perhaps dominating at even larger masses.

\end{abstract}


%% ----------------------------------------------------------------------------------------------------------------------------------
\section{Introduction}\label{intro}

Hierarchical growth, by which increasingly larger structures are built through the assembly of smaller ones, is a major
feature of the $\Lambda$CDM paradigm.  Its imprint on the evolving abundance of galaxy clusters is an important
cosmological probe \citep[e.g.,][]{vikhlinin09} and evidence for hierarchical growth has also been reported among
group-scale halos \citep{williams12}.  Because galaxies reside in dark matter halos, one also expects patterns of
hierarchical growth in observables that trace galaxy mass such as stellar mass, $M_*$ \citep{stringer09}, or assembly
history \citep[e.g.,][]{gu16}, including morphology \citep{wilman13} and size \citep{zhao15}.

Indeed, recent galaxy formation models employing both hydrodynamic simulations and semi-analytic recipes predict a
galaxy stellar mass function that grows substantially at the high-mass end, tracking to some degree the dark matter
halo mass function \citep[e.g.,][]{de-lucia07,guo11,furlong15,torrey17}.  While various but still uncertain mechanisms
limit star formation among both low- and high-mass galaxies \citep{benson03}, thus working to decouple $M_*$ from
$M_{\rm halo}$, late-time growth in $M_*$ among the most massive galaxies (with no ongoing star formation) is still
expected as a result of galaxy mergers \citep[e.g.,][]{lee13,qu17}.

The role of such mergers in driving high-mass galaxy growth at $z \lesssim 2$ has been the subject of recent
observational work \citep[e.g.,][]{bundy09,lotz11,casteels14, mundy17} and the basis of theoretical explanations for
how massive compact spheroidals at $z \approx 2$ grow significantly in size by the present day
\citep[e.g.,][]{hopkins10a, nipoti12, hilz13, welker17}.  Comparisons of the predicted growth in diffuse outer
components required to drive increasing size estimates appear to be consistent with observed (minor) merger rates, at
least for $z \lesssim 1$ \citep{newman12, lopez-sanjuan12, ownsworth14}.

The rate of merging required to grow high-mass galaxies sufficiently in size should also add significantly to their
stellar mass \citep[e.g.,][]{lidman13}.  An implied $\sim$30\% growth in $M_*$ since $z \sim 1$ is typical and should
be reflected in derived $M_*$ growth rates from evolving galaxy stellar mass functions.  Recent observational results,
however, have largely indicated little or no evolution in the total mass function and a lack of $M_*$ growth from $z
\sim 1$ to today \citep [e.g.,][]{brammer11, moustakas13, ilbert13, muzzin13, davidzon13}.  How can hierarchical
assembly explain the growth in galaxy sizes but not simultaneously yield growth in galaxy masses?

One answer is that we are only beginning to survey the large volumes required to detect the expected signal.
\citet{stringer09} argue that tens, if not hundreds, of deg$^2$ are required to statistically confirm hierarchical
growth in galaxy mass functions.  In this regime, attention to systematic uncertainties is critical \citep[e.g.,][]{marchesini09}. Much recent work on galaxy number densities has prioritized reaching higher redshifts
with ``pencil-beam'' surveys that sample combined areas of only a few deg$^2$.  \citet{moustakas13}, which is based on
the 5.5 deg$^2$ PRIMUS survey \citep[PRIsm MUlti-object Survey,][]{coil11} and \citet{davidzon13}, which analyzes early
data obtained over 10.3 deg$^2$ from the VIPERS survey \citep[The VIMOS Public Extragalactic Redshift
Survey,][]{guzzo14}, represent early attempts to extend $M_*$-complete redshift surveys to larger areas.

% brammer11: Newfirm MBS; 
% ilbert13: 1.5 deg^2 in one field
% muzzin13: also Ultra-VISTA
% tomczak14: 316 arcmin^2 plus NEWFIRM 
% davidzon13 : 10.3 deg^2
% moutard16: 22.4 deg^2

To reach larger cosmic volumes, the challenge of building complete spectroscopic samples makes photometric redshifts
(\photozs{}) attractive, especially as wide-and-deep imaging surveys become more prevalent.  \citet{moutard16} exploit
VIPERS PDR-1 \citep{garilli14} spectroscopic redshifts (\speczs{}), CFHT, and GALEX photometry obtained over the VIPERS
footprint to construct a \photoz{}-based galaxy sample with $0.2 < z < 1.5$ that is complete to $\approx 10^{10} \msun$
at $z=1$.  This sample is used to study the evolving mass function over 22 deg$^2$ in \citet{moutard16a}.  Some years
earlier, \citet{matsuoka10} combined and reanalyzed imaging data from the SDSS Stripe 82 Coadd \citep[see][]{annis14}
and the UKIDSS Large Area Survey \citep[LAS,][]{lawrence07} in order to derive photometric redshifts and study galaxy
mass functions over 55 deg$^2$ at $z < 1$.  The analysis we present in this work utilizes these same data sets, which
have become more complete since \citet{matsuoka10} and can be combined with a substantial number of \speczs{} to yield
an $M_*$-complete sample comprising 139 deg$^2$ with $z < 0.7$, part of what we term the Stripe 82 Massive Galaxy
Catalog \citep[the \cat{},][]{bundy15a}.

With tens of square degrees surveyed, both \citet{matsuoka10} and \citet{moutard16} claim to detect growth in the
number density of the most massive galaxies although the amplitude of detected evolution is inconsistent.  At $\log
M_*/\msun > 11.5$, \citet{matsuoka10} find nearly an order of magnitude increase in number density from $z \sim 1$ to
$z \sim 0.3$ while \citet{moutard16} measure only a factor of 2 increase.  Meanwhile, initial work by \citet{capozzi17}
exploits 155 deg$^2$ of the Dark Energy Survey (DES) Science Verification Data to report a modest \emph{decrease} in
$M_*$ at the highest masses since $z \approx 1$.

This discrepancy highlights the challenge of this measurement and raises concerns about uncertain (perhaps
catastrophically uncertain) \photozs{} as well as possibly larger-than-expected contributions from ``cosmic variance.''
Both issues can be addressed by turning to very wide \specz{} surveys designed to constrain cosmological parameters via
angular clustering.  The drawback of these surveys, which can span thousands of deg$^2$, is the difficulty accounting
for incompleteness owing to the selection criteria.  Relevant here is early work by \citet{wake06} that detected no
evolution in the number density of ``luminous red galaxies'' as measured at $z \sim 0.55$ by the 2SLAQ survey and at $z
\sim 0.2$ by the Sloan Digital Sky Survey (SDSS).  Finding a consistent luminosity function as that measured in the
magnitude-limited COMBO17 survey, \citet{wake06} argue that the no evolution conclusion applies broadly to the
high-mass galaxy population.

\citet{maraston13} present a more recent example of this approach, using \speczs{} from the SDSS-III BOSS survey
\citep[Baryon Oscillation Spectroscopic Survey,][]{dawson13} taken from the $0.43 < z < 0.7$ CMASS (``constant mass'')
sample.  Instead of correcting for incompleteness in the CMASS sample, \citet{maraston13} apply the same CMASS
selection cuts to simulated data from a semi-analytic model. Doing so indicates that at least for $z \lesssim 0.6$,
CMASS reaches high completeness ($>$90\%) at the highest masses (a conclusion that is confirmed and quantified in
\citealt{leauthaud16}).  In agreement with the earlier \citet{wake06} result, the CMASS mass function at the highest
masses shows no evolution over $0.45 < z < 0.7$.  The \citet{maraston13} analysis is based on 283,819 galaxies spanning
3275 deg$^2$.

The question of whether the total mass function evolves has implications for the separate evolution in the numbers of
star-forming and passive galaxies.  At masses below 10$^{11} \msun$, there is broad agreement that the number of
``quenched'' galaxies increases with time \citep[e.g.,][]{bundy06,borch06,drory09,ilbert10,moustakas13}, but some
controversy remains over whether the star-forming population remains constant \citep[e.g.,][]{ilbert10,moutard16a} or declines
\citep[e.g.,][]{moustakas13}, especially at $M_* > 10^{11}$.  A constraint from the total mass function would help
distinguish the extent to which star-formers shut down and transform into quenched galaxies versus the rate of new
arrivals (from lower $M_*$) that either replenish the star-forming population \citep[e.g.,][]{peng10} or add to the
increasing number of quiescent galaxies.

% where uncertainties are large \citep[e.g.,][]{marchesini09}.

The purpose of this work is to study number density evolution at the highest masses using a sample that combines
well-understood completeness functions typical of magnitude-limited surveys with large spectroscopic datasets designed
to constrain cosmological parameters.  In \citet{bundy15a} (hereafter \catpaper{}), we build such a sample by combining
SDSS Coadd $ugriz$ photometry in the ``Stripe 82'' region \citep{annis14}, reaching $r$-band magnitudes of $\sim$23.5
AB, and near-IR photometry in $YJHK$ bands to 20$^{\rm th}$ magnitude (AB) from the UK Infrared Deep Sky Survey Large
Area Survey \citep[UKIDSS-LAS,][]{lawrence07} with 70,000 \speczs{} from the SDSS-I/II and the Baryon
Oscillation Spectroscopic Survey.  We refer to the combined data set as the Stripe 82 Massive Galaxy Catalog (\cat{})
and make it publicly available at {\href{http://www.massivegalaxies.com}{MassiveGalaxies.com}}.  \comppaper{} in this series, \citet{leauthaud16},
uses the \cat{} to investigate the $M_*$ completeness limits of BOSS \specz{} samples.  The \cat{} was also used in
\citet{saito16} to constrain the relationship between high-mass galaxies and their dark matter halos.  In this paper,
\mfnpaper{}, we use an $M_*$-complete sub-sample of the \cat{}, comprising 139 deg$^2$ and sampling \vol{} Gpc$^3$, to
measure galaxy mass functions with unprecedented precision at $\log M_*/\msun > 11.3$ over $0.3 < z < 0.65$.  Finding
no apparent evolution, we place particular emphasis on how scatter in $M_*$ measurements, biases resulting from
assumptions underlying $M_*$ estimates, and other uncertainties limit the interpretation of our results.

A plan of the paper is as follows.  We begin in Section \ref{data} by summarizing the key components of the \cat{} and
its construction.  Full details can be found in \catpaper{}.  The various $M_*$ estimates used in this work are
described in Section \ref{mstar}.  We discuss potential biases in derived mass functions for large samples including
the impact of various forms of measurement scatter in Section \ref{sec:methods}.  Our results are presented in Section
\ref{results}, where we study how the adoption of different priors (Section \ref{MF:priors}) and stellar population
synthesis models (Section \ref{MF:models}) affect the degree of evolution we infer.  The mass functions of galaxies
with different levels of residual star formation are presented in Section \ref{MF:SFH} and made available at \href{http://www.massivegalaxies.com}{MassiveGalaxies.com}.  We discuss the significance of
our results and their limitations as well as comparisons to other work in Section \ref{discussion}.  Section
\ref{summary} provides a summary.  Throughout this paper, we use the AB magnitude system and adopt a standard cosmology
with $H_0$=70 $h_{70}$ km s$^{-1}$ Mpc$^{-1}$, $\Omega_M$=0.3 and $\Omega_{\Lambda}$=0.7.

%% ----------------------------------------------------------------------------------------------------------------------------------
\section{The Stripe 82 Massive Galaxy Catalog}\label{data}

Full details of the \cat{} catalog construction are presented in \catpaper{}.  We summarize key aspects here with a
focus on the final \ukwide{} sample that we use for our mass function analysis.

\subsection{$ugrizYJHK$ Photometry}\label{photometry}

The ``SDSS Coadd'' provides the primary source catalog for the \cat{}.  This data set refers to repeated $ugriz$
imaging in Stripe 82 (-50\arcdeg{} $<$ $\alpha_{\rm J2000}$ $<$ +60\arcdeg{}) first presented in \citet{abazajian09}
and further described in \citet{annis14}.  The point-source 50\% completeness limit for the Coadd is $r \sim 24.4$
(AB). The Coadd photometric catalog is queried as described in \catpaper{} to define a unique sample which is then
cross matched to overlapping near-IR data from the Large Area Survey (LAS) component of UKIDSS \citep{lawrence07} Data
Release 8 (DR8).  The LAS aims to reach AB magnitude depths of $Y = 20.9$, $J=20.4$, $H=20.0$, and $K=20.1$, but we
provide field-dependent measures of the achieved depth in the \cat{} and use these to define an areal footprint that
satisfies specific depth requirements in the \ukwide{} selection described below.

PSF-matched $ugrizYJHK$ photometry in the \cat{} is obtained with the \synmag{} software \citep{bundy12} which uses
SDSS surface brightness profile fits to predict the SDSS $r$-band magnitude that would have been obtained using the
same aperture and under the same atmospheric seeing as magnitudes measured in each UKIDSS filter.  For total $H$- and
$K$-band magnitudes, which form the basis of our $M_*$ estimates, we overcome biases resulting from blended sources in
the UKIDSS photometry by building a new flux estimator referenced to the SDSS $z$-band \texttt{CModelMag} magnitude.
After correcting for the aperture-matched optical-to-near-IR color (e.g., $(z-K)$), we define \texttt{HallTot}
magnitudes by adjusting the reported Hall magnitudes to match \texttt{CModelMag$_z$} on average.  For blended sources,
which are known to have biased Hall magnitudes, we set the \texttt{HallTot} magnitude to \texttt{CModelMag$_z$} and
apply the color correction.  Further details are given in \catpaper{}.

\subsection{Spectroscopic and Photometric Redshifts}
% PLOT:  Photo-z vs. Spec-z for RedMapper and ANNz
%   Made in boss/mkbest.pro with /PLOT
\begin{figure*}
\epsscale{1.2}
\plotone{\dir_plots/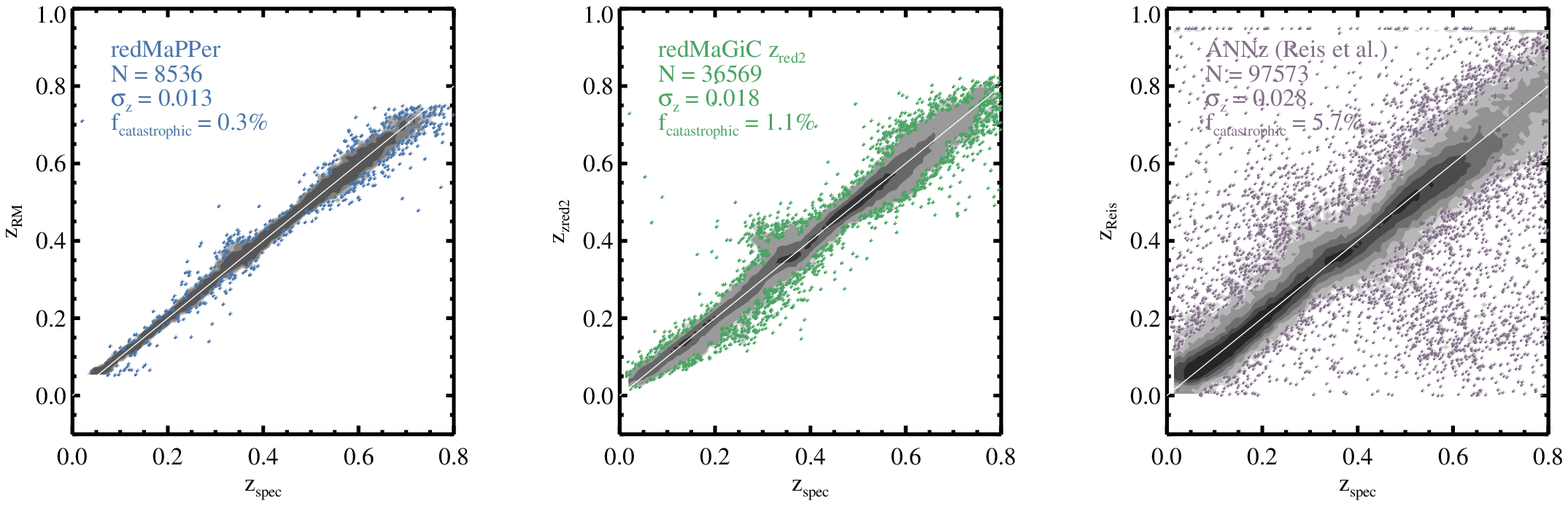}
% \centerline{\includegraphics[scale=0.8]{maghist_remeasure.pdf}}
\caption{Comparisons of three photometric redshift estimators to available spectroscopic redshifts in Stripe 82,
reproduced from \catpaper{}.  The
  comparison is limited to $i < 22.5$ and $0.01 < z_{\rm spec} < 0.8$.  The left and middle panels are from the
  \redmapper{} project \citep{rozo15}, while the right panel compares neural-network \photozs{} from
  \citet{reis12}. The 3$\sigma$-clipped dispersion is listed in each panel along with the fraction of catastrophic
  outliers defined by $\abs{\Delta z} > 0.1$.  Contours are plotted at high data densities with 0.3 dex logarithmic
  spacing in the left and middle panel and 0.4 dex in the right panel.  The 1-to-1 relation is plotted in each panel as
  a thin light grey line.
  \label{fig:photoz_specz}}
\end{figure*}

The SDSS-III program \citep[][]{eisenstein11} BOSS program provides 149,439 spectroscopic redshifts for the \cat {}.
Redshifts from the LOWZ, CMASS, and ``Legacy'' samples, as collated in the SDSS-III SpecObj-dr10 catalog, are all
included.  We combine photometric redshifts from a number of sources to supplement the \cat{} when \speczs{} are not
available.  For the bright galaxies we study in this work ($i \lesssim 22.5$), we define \zbest{} to be the
spectroscopic measurement, if available.  If a \photoz{} is required, we first check if the galaxy resides in a cluster
with a redshift assigned by the red-sequence Matched-filter Probabilistic Percolation
\citep[\redmapper{},][]{rykoff14}.  Defining $\sigma_z$ as the 3$\sigma$-clipped standard deviation of $\Delta z =
z_{\rm spec} - z_ {\rm phot}$ (note that we do not divide by $1+z$) and catastrophic outliers as those with
$\abs{\Delta z} > 0.1$, 
the \redmapper{} \photozs{} have $\sigma_z \sim 0.02$ and a catastrophic rate of less than 1\%.  For
field galaxies on the red sequence, we adopt estimates from the red-sequence Matched filter Galaxy Catalog
\citep[\redmagic{},][]{rozo15}.  These are only slightly worse in terms of \photoz{} quality.  If neither the
\redmapper{} nor \redmagic{} \photozs{} are available, we assign \zbest{} to the neural network results derived in
\citet{reis12}.  The \citet{reis12} redshifts have $\sigma_z \sim 0.03$ and a 5\% outlier fraction at $z \sim 0.5$.
Comparisons of these three \photoz{} estimators to available \speczs{} are presented in Figure \ref{fig:photoz_specz}
and refer the reader to \catpaper{} for further discussion of redshift reliability and completeness.

At $\log{M_*/\msun} > 11.4$ and $z \sim 0.6$, the \ukwide{} sample we define below has a \specz{} completeness of 80\%.
 Of the remaining galaxies without \speczs{}, $\sim$8\% have \redmagic{} \photozs{}.  A roughly equal number have
 \citet{reis12} \photozs{}, and a few percent come from \redmapper{}.  The \specz{} completeness improves towards lower
 redshifts and higher $M_*$ (\comppaper{}).  We also note that \citet{pforr13} found little bias ($\sim$0.02 dex) when
 comparing $M_*$ estimates based on \photozs{} compared to \speczs{} for passive galaxies.

\subsection{The \ukwide{} sample}\label{sec:ukwide}\label{sample}

The mass functions discussed below are derived using a subset of 517,714 galaxies in the \cat{} called the \ukwide{}
sample.  The selection criteria are described in detail in \catpaper{} and include star-galaxy separation, the
application of rejection masks in all bands, photometry quality flags, and 5$\sigma$ $YJHK$ imaging
depths of $[20.32, 19.99, 19.56, 19.41]$ in AB magnitudes.  The resulting \ukwide{} sample spans \area{} deg$^2$,
and is complete above $\log M_*/\msun \approx 11.3$ at $z = 0.7$.

%% ----------------------------------------------------------------------------------------------------------------------------------
\section{Stellar Mass Estimates}\label{mstar}

As we show in Section \ref{results}, systematic uncertainties in $M_*$ estimates dominate conclusions about high mass
galaxy growth in the \cat{} sample.  In this section, we present a set of $M_*$ estimates based on the same photometric
data set, and study systematic offsets that arise when different priors, models, and variants of the photometry are
used.  In Section \ref{results}, we will show how $M_*$ offsets translate into systematics in the recovered stellar
mass function.  For comparisons with publicly available BOSS $M_*$ estimates\footnote{\citet{tinker17} suggest that the
``Wisconsin PCA'' $M_*$ estimates have the smallest measurement uncertainties among available BOSS estimates.  While
they are compared in \catpaper{}, we do not use them here because they are available only for galaxies with
spectroscopic redshifts.}, please see \catpaper{}.

\subsection{The \cat{} Fiducial $M_*$ Estimates}

We recount the description of the \cat{} $M_*$ estimates presented in \catpaper{}.  These fiducial $M_*$ estimates (we
will distinguish them with the label, \Mmgc{}) are derived using the Bayesian code developed for mass function work in
\citet{bundy06} and \citet{bundy10}.  The observed SED of each galaxy is compared to a grid of 13440 \citet{bruzual03}
(BC03) population synthesis models, including 16 fixed age values and 35 fixed exponential timescales, $\tau$.  Ages
are drawn randomly from a uniform distribution between 0 and 10 Gyr and are restricted to less than the cosmic age at
each redshift. Values for $\tau$ are also random in the linear range between 0.01 and 10 Gyr.  No bursts are included
and the dust prescription follows \citet{charlot00}. See Table \ref{tab:fits}.  We assume a Chabrier IMF
\citep{chabrier03}, $\Omega_M$=0.3, $\Omega_{\Lambda}$=0.7, and a Hubble constant of 70 km s$^{-1}$ Mpc$^{-1}$.

At each grid point, the \kband\ $M_*/L_K$ ratios (corresponding to the ``current'' mass in stars and stellar remnants),
inferred $M_*$, and probability that the model matches the observed SED is stored.  This probability is marginalized
over the grid, giving an estimate of the stellar mass probability distribution\footnote{Note that we assume the prior
grid adequately samples the parameter space of the posterior.}.  We take the median as the final estimate of $M_*$. The
68\% width of the distribution provides an uncertainty value which is typically $\sim$0.1 dex.

\subsection{$M_*$ Estimates from \isedfit{}}

We also produce $M_*$ estimates (\Mised{}) using the Bayesian \isedfit{} package presented in \citet{moustakas13}.  The
\isedfit{} code has several advantages.  In addition to performing a refined grid search of the $M_*$ posterior
distribution and enabling priors with non-flat probability distributions, \isedfit{} can return $M_*$ estimates for
multiple stellar population synethsis models, including FSPS \citep{conroy09}, BC03 \citep{bruzual03}, and Maraston
\citep{maraston05} models.

The basic set of \isedfit{} priors is similar to those used for the \Mmgc{} estimates and are based on a set (randomly
generated for each run of \isedfit{}) of 25000 declining exponential models.  The \Mised{} estimates additionally
include a prescription for bursts described below.  Unlike the \Mmgc{} models whose parameters fall on a grid, the
parameters for each \isedfit{} model vary independently, better sampling the range of each prior.  The \isedfit{} ages
are restricted by the cosmic age at each redshift and drawn linearly from the range, 0.1--13 Gyr.  The exponential
$\tau$ prior is drawn from the linear range, 0.1--5 Gyr.  The metallicity and dust assumptions are similar to the
\Mmgc{} estimates. The \isedfit{} code is designed to work with flux measurements which we take directly from a
conversion of SDSS ``Luptitudes'' for $ugriz$ and via a transformation to AB magnitudes for the UKIDSS photometry.

In the case of the \Mised{} fits, stochastic bursts are added randomly to the star formation histories.  For every 2
Gyr interval over the lifetime of a given model, the cumulative probability that a burst occurs is 0.2.  Each bursts'
SFH is Gaussian in time with an amplitude set by, $\mathcal{F}_b$, the total amount of stellar mass formed in the burst
divided by the underlying mass of the smooth SFH at the burst's peak time.  $\mathcal{F}_b$ is drawn from the range
0.03--4.0.  The allowed burst duration ranges from 0.03--0.3 Gyr.

Table \ref{tab:fits} lists several \isedfit{} runs we have performed.  The impact of the resulting $M_*$ estimates on the derived mass function is discussed in Sections \ref{MF:priors} and
\ref{MF:models}.

	% Mstar estimator Table: 
		\begin{deluxetable*}{lcccc}
		\tablecaption{$M_*$ estimators}
		\tabletypesize{\footnotesize}
		\tablewidth{0pt}
		\tablecolumns{5}
		\tablehead{
		\colhead{name} & \colhead{models} & \colhead{main priors} & \colhead{bursts} & \colhead{$M_*$ scaling} \\
		%\cline{1-12} \\
		%\multicolumn{12}{c}{EGS Field}
		}

		\startdata

		\Mmgc{} 		& BC03	& \citet{bundy06} & none & \kband{} \\[0.1cm]
		\MmgcOpt{} 		& BC03	& \citet{bundy06} & none & $z$-band \\[0.1cm]
		\Mised{} 		& FSPS 	& PRIMUS \citep{moustakas13} & $P_{\rm burst}=0.2$ & average \\[0.1cm]
		\MisedFSPS{} 	& FSPS 	& PRIMUS \citep{moustakas13} & none & average \\[0.1cm]
		\MisedBC{} 		& BC03 	& PRIMUS \citep{moustakas13} & none & average \\[0.1cm]
		\MisedMa{}	 	& Maraston 	& PRIMUS \citep{moustakas13} & none & average \\

		\enddata
		\label{tab:fits}
		%\tablecomments{}
		\end{deluxetable*}

\subsection{Optical vs. Near-IR Photometry}

        % PLOT:  mass comparison: Optical versus mass, redshift, CModelMag_i, B1000
          %   Made in boss/masscomp.pro with /OPTPLOT, Chapter_31
          \begin{figure*}
          \epsscale{1.2}
          \plotone{\dir_plots/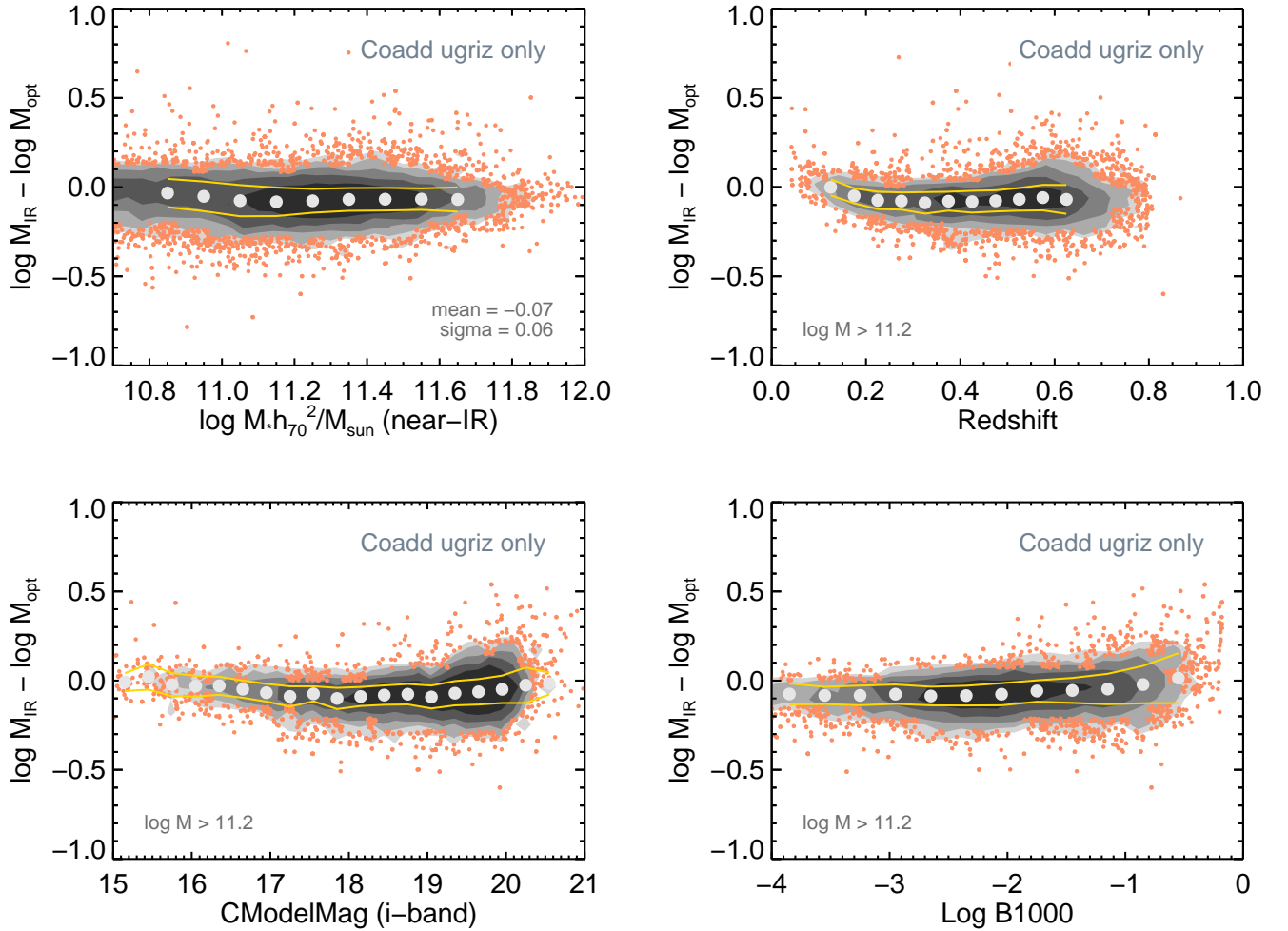}

    \caption{Comparison of \Mmgc{} estimates from SED fitting applied to optical photometry only ($ugriz$) with those
    from optical+near-IR photometry ($ugrizYJHK$).  For near-IR masses, $M_*$ is estimated by scaling the determined
    $M/L$ in the observed \kband{} by the observed luminosity in that band as measured using the \texttt{KHallTot}
    magnitude. For optical masses, the observed-frame $z$-band $M/L$ is scaled by SDSS Coadd $z$-band \texttt
    {CModelMag}.  The optical-near-IR $M_*$ difference is plotted as a function of near-IR $M_*$ (top left), redshift
    (top right), $i$-band \texttt {CModelMag} (bottom left), and the birth parameter, \birth{}, a measure of the
    inferred SFR averaged over the last 1000 Myr compared to the SFR averaged over the galaxy's lifetime.
    \label{fig:masscomp_opt}}

          \end{figure*}

Providing photometric coverage in the near-IR, which is more sensitive to older stellar populations that typically
dominate $M_*$, was one of the motivations for assembling the \cat\ \citep{bundy15a}.  We can test the impact of
near-IR photometry by comparing the standard \Mmgc\ estimates, which are based on $ugrizYJHK$, to those using solely
the $ugriz$ bands (we label these \MmgcOpt).  We use the \Mmgc\ mass estimator in both cases.  Figure
\ref{fig:masscomp_opt} tracks the mass difference, \Mmgc$ - $\MmgcOpt, as a function of several parameters.  For
masses above $\Mmgc > 10^{9} \msun$ the top-left panel reveals a small offset of -0.07 dex with a scatter of 0.06 dex
but no strong dependencies on \Mmgc.  The difference in mass estimates systematically changes for lower redshift
galaxies with apparent magnitudes brighter than $\sim$17 AB (top-right and bottom-left panels).  The final
panel in Figure \ref{fig:masscomp_opt} investigates the dependence on \birth{}, a measure of recent star formation composed of the ratio of the star formation rate (SFR) averaged over the last 1000 Myr to the average SFR over the galaxy's lifetime.  This panel shows that across mass and redshift, galaxies in the \cat\ with higher
\birth\ values, implying more recent star formation, deviate from the -0.07 dex offset that defines \Mmgc$ - $\MmgcOpt{} for most of the sample.  These galaxies show offsets that are $\sim$0.1 dex larger and have $\log M_*/\msun \lesssim 11.5$.

The systematic differences in $M_*$ that are evident in Figure \ref{fig:masscomp_opt} arise from two sources.  First,
near-IR photometry provides additional constraints on galaxy SEDs that should yield better estimates of mass-to-light
(M/L) ratios.  Errors from photometric matching across many bands could also degrade the SED fit quality, however.
Figure \ref {fig:masscomp_opt} shows that for the redshifts relevant to this work ($z > 0.2$), including near-IR
constraints has little or no effect on $M_*$ estimates, suggesting that $ugriz$ photometry alone provides similar
estimates for massive galaxies at $z < 0.8$ as does optical plus near-IR photometry.  Perhaps not surprisingly,
however, the role of near-IR data becomes important for the modest number of massive galaxies with recent star
formation (bottom-right panel).  Here, assuming that the near-IR masses are more accurate, the optical-only estimates
may be biased low by -0.1 dex with deviations as high as -0.5 dex in individual cases.

The second factor behind systematic differences in Figure \ref{fig:masscomp_opt} is the use of different total flux
estimators.  The \Mmgc\ estimates are the result of multiplying the M/L derived for the observed-frame
$K$-band\footnote{In rare cases, the $H$-band is used when $K$-band is not available.} by 
\texttt{KHallTot}, a non-parametric total magnitude estimate.  As discussed \catpaper{}, the \texttt{KHallTot}
measurements are less biased by blended sources compared to other flux estimators in the
UKIDSS photometry.  However, \texttt{KHallTot} must be adjusted globally to match the $z$-band \texttt
{CModelMag} estimates.  The \MmgcOpt\ estimate, on the other hand, is the direct
product of the observed-frame $z$-band M/L and the $z$-band \texttt{CModelMag}.  The \texttt{CModelMag} estimator
combines total flux measures from SDSS-derived, 2D fits of an exponential and a de Vaucouleurs surface brightness
profile.  Differences in the way \texttt{CModelMag} and \texttt{HallTot} account for the ``total light'' in a surface
brightness profile can therefore impact the $M_*$ measurements.

        % PLOT:  magnitude comparison: CModel versus KHallTot and Coadd vs. DR10
          %   Made in boss/mag.pro with /MAG1, Chapter_31
          \begin{figure*}
          \epsscale{1.2}
          \plotone{\dir_plots/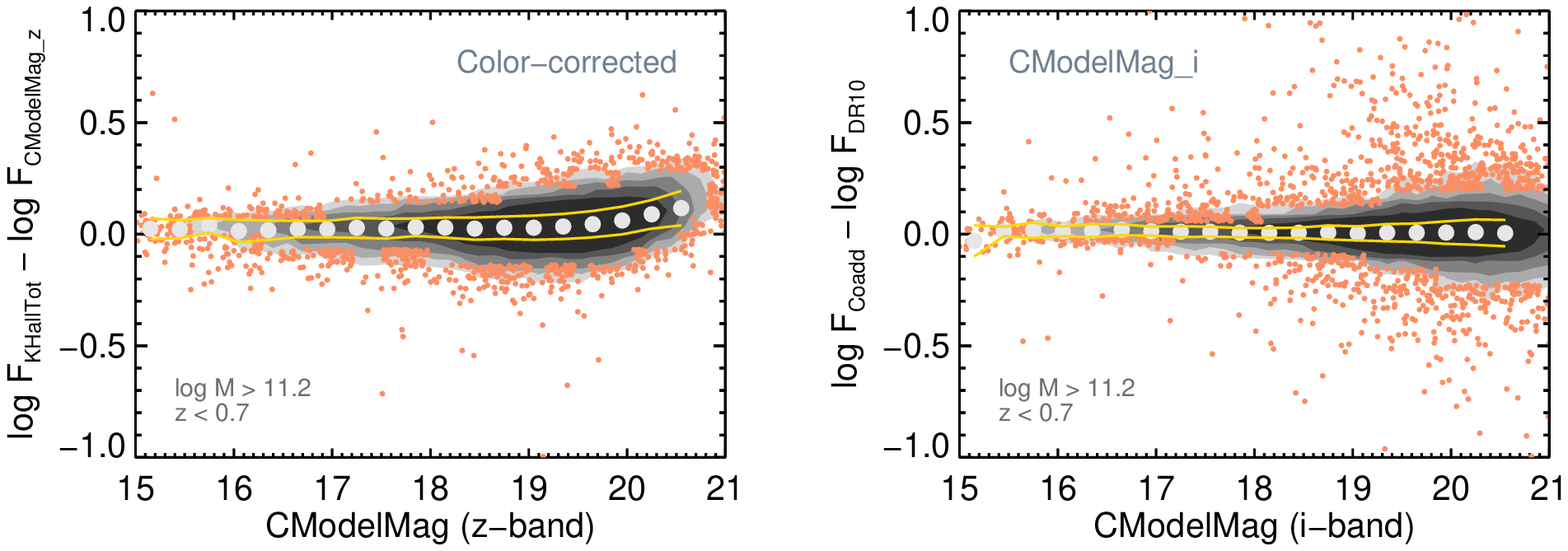}

    		\caption{Comparisons of total flux estimators relevant to scaling $M_*$ estimates.  The left panel compares
    		the \kband{} flux from \texttt{KHallTot} (used in \Mmgc{}) to the $z$-band \texttt{CModelMag} flux
    		\emph{after} correcting for the aperture-matched color difference between the bands ($z-K$).  A similar
    		behavior is seen in the bottom-left panel of Figure \ref{fig:masscomp_opt}, indicating that much of the
    		difference in (\Mmgc
    		$-$
    		\MmgcOpt) is driven by the flux estimator.  The right-hand panel shows consistency in $i$-band
    		\texttt{CModelMag} measurements between the Coadd photometry and single-epoch DR10 ($z$-band shows similar
    		behavior). \label{fig:magcomp1}}

          \end{figure*}

Figure \ref{fig:magcomp1} explores this by comparing the flux corresponding to \texttt{KHallTot} to that from the
$z$-band \texttt{CModelMag}$_z$ as a function of \texttt{CModelMag}$_z$ (left panel).  The effect of $(z-K)$ color
(derived from PSF-matched photometry) has been removed.  The flux difference remains flat until \texttt{CModelMag}$_z$
$\sim$ 19 AB, at which point the near-IR flux estimator grows slightly in comparison to \texttt{CModelMag}$_z$.
\texttt{KHallTot} is 0.1-0.2 dex brighter at the faintest magnitudes in the sample.  This trend seems to be expressed
in the direct $M_*$ comparison against \texttt{CModelMag}$_i$ (Figure \ref{fig:masscomp_opt}, bottom-left).  But
systematic differences in total flux cannot explain the increasing $M_*$ discrepancy at \texttt{CModelMag}$_i < 17$~AB
and, correspondingly, $z < 0.2$.  In this very bright regime, changes in the M/L inferred from SED fits to
the multi-band photometry must be responsible.  We do not pursue these $M_*$ offsets further because, for the mass
function analysis that follows, we restrict ourselves to higher redshifts.  

A final test is provided in the right-hand panel of Figure \ref{fig:magcomp1} which compares \texttt{CModelMag}
estimates from the SDSS Coadd (used in the \cat) to those from the single-epoch SDSS photometry.  There is an expected
increase in scatter at fainter magnitudes (because the Coadd is much deeper) but no evidence for systematic trends.
Given that \texttt{CModelMag}$_i$ is the dominant, and often sole, total magnitude used to normalize other $M_*$
estimates provided by the BOSS team, the good agreement shown here makes the Coadd-based \cat\ a valuable anchor
for understanding $M_*$ systematics in studies exploiting the full BOSS data set \citep[e.g.,][]{maraston13}.

%% ----------------------------------------------------------------------------------------------------------------------------------
\section{Methods: Number density distributions in large-volume surveys}\label{sec:methods}

One of the goals of this paper is to use the \cat{} to explore the new ``large-volume'' regime for \emph{complete}
studies of galaxy number density distributions, such as the mass and luminosity functions.  For samples spanning more
than $\sim$100 deg$^2$ and a significant redshift baseline, several considerations arise.  An obvious point is that
making use of the statistical precision afforded by large volumes requires careful control of the error budget. 
Ideally, we would restrict ourselves to using only spectroscopic redshifts for this reason, but in the near term,
obtaining \speczs{} for the tens of millions of sources that current imaging surveys now detect (e.g., to $i < 23$ AB)is infeasible.  

Even if we limit ourselves (as we do in the next section) to brighter subsamples where spectroscopic follow-up is
possible, we are left with the challenge of determining the completeness of the sample at a level of precision on par
with that of the number density measurements themselves.  One must either additionally commit significant spectroscopic
resources to defining the completeness limit (i.e., ``throwing away'' a large number of hard-earned \speczs{}),
estimate the completeness by applying the selection criteria to simulated samples \citep[see][]{maraston13}, or turn to
photometric redshifts as we do in this work to supplement redshift information where the \speczs{} are incomplete.  In
the sample we use below, the fraction of galaxies with $\log M_*/\msun > 11.4$ (Chabrier IMF) and $z < 0.6$ that
require \photozs{} is roughly 20\% \citep{leauthaud16}.

The introduction of \photozs{} adds sources of both random and systematic error that must be accounted for
 \citep[e.g.,][]{etherington17}. At the same time, a new tool for diagnosing such errors becomes available when the
 expected, random statistical fluctuations (including sample variance) are negligible.  That tool is essentially a
 prior that dictates that the shape and normalization of actual number density distributions should evolve smoothly
 with redshift.  A stronger version would assume that evolution in the average properties of the galaxy distribution is
 both smooth and monotonic. A particular redshift bin, for example, with a clear excess number density can be a sign
 post of systematic errors that preferentially affect those redshifts.

\subsection{Biases from Photometric Redshifts}

While the mass functions derived here rely on a sample with $\gtrsim$80\% \specz{} completeness, the use of \photozs{} can introduce errors in a number of ways.  These include biases in the binned redshift distribution itself, scatter in the luminosity distance used to normalize $M_*$, and errors on the recovered rest-frame SED.  

To first order, \photozs{} introduce a Gaussian redshift uncertainty, blurring out structure in the true redshift
distribution and creating contamination between adjacent redshift bins.  The effect of contamination is reduced as the
bin size increases above the 1-$\sigma$ \photoz{} uncertainties.  If the \photoz{} uncertainty depends on redshift, the
bin-to-bin contamination will vary with $z$ as well.  Defined $z$-bins at the limits of the full range accessible will
also have true redshift distributions that are asymmetric.  These effects are typically small because \photoz{}
uncertainties of $\sigma_z \sim 0.03$--0.07 can often be achieved and are usually smaller than the redshift baselines
probed ($\Delta z > 0.3$).  These uncertainties also depend weakly on $z$ across most samples.  Finally, biases in the
mean \photoz{} are often much smaller than $\sigma_z$.

The larger impact of roughly Gaussian \photoz{} uncertainties is the contribution of an additional random error on the
 $M_*$ (or $L$) estimates as a result of their dependence on the now-uncertain luminosity distance.  The resulting \photoz{}-induced scatter in $\log M_*$, which we refer
 to as $\sigma_{M_*,z}$, as a result of the \photoz{} uncertainty, $\sigma_z$, can be estimated as $\sigma_{M_*,z} \approx
 \sigma_z/z$.  Among the worst \photozs{} ($\sigma_z = 0.04$) in the \cat{} at $z \approx 0.6$, for example, the
 \photoz{} uncertainty adds 0.07 dex in quadrature to the $M_*$ errors, which exhibit $\sigma_{M_*} = 0.1$--$0.2$ dex
 when $z$ is perfectly known.  We will discuss
 how random errors in $M_*$ can be addressed in the next section. 

In addition to making luminosity distances more uncertain, the \photoz{} error shifts the inferred rest-frame wavelength of the SED, thereby degrading the quality of the fit and derived M/L ratio.  This effect is small.  Tests applied to the \cat{} neural-network \photozs{}, which have $\sigma_z \approx 0.02-0.03$, indicate that this restframe color uncertainty adds 0.02 dex in quadrature to $\sigma_{M_*}$.

A more subtle but extremely important problem occurs when the \photoz{} scatter increases over a specific range in
redshift.  A look at the \photoz{}-\specz{} comparison in Figure \ref{fig:photoz_specz} shows an often-seen
degradation in \photoz{} quality at $z \approx 0.35$ that corresponds to the 4000 \AA{} break falling between the $g$
and $r$ bands.  With only a few thousand \speczs{} to compare against, this feature is hardly noticeable.  Furthermore,
because the additional scatter appears roughly symmetric, it is tempting to believe that any
effect on the mass or luminosity function would cancel out.  

When tens of thousands of \speczs{} are available, as in the \cat{}, this \photoz{} feature reveals itself to be much
more prominent, with a noticeable tail.  The key point is that the \emph{direction} of \photoz{}
scatter can have a profound impact on derived number density functions.  Up-scattering yields a greater
distance for a galaxy, shifting it into a higher \photoz{} bin and assigning a higher $M_*$ or $L$ than it deserves.
Because more massive and intrinsically luminous galaxies are significantly rarer than their low-mass counterparts,
up-scattering can create a significant bias in the reported number density evolution.  Even when \photoz{}
down-scattering is symmetric, it has a less significant impact because the true number of lower-mass galaxies significantly outweighs the number of contaminants.

Similar arguments apply to the location of catastrophic \photoz{} outliers.  For these and other kinds of \photoz{}
behavior, it may be possible to influence \photoz{} codes so that they fail in preferred ways.  In others, the choice
of redshift bins can be designed to avoid regions of worrisome contamination.  It may also be possible to model
\photoz{} effects and account for them, although this is beyond the scope of the current paper.

\subsection{Accounting for scatter in $M_*$ or $L$}\label{method:scatter}

Even with \specz-only samples, random errors in the $M_*$ (or $L$) estimates introduce Eddington bias in the derived
galaxy mass functions as a result of the steep decline in the number of galaxies at the bright end.  The contamination
from intrinsically lower-mass galaxies scattering upwards outweighs the down-scattering of higher mass galaxies because
there are many more lower-mass galaxies subject to random $M_*$ errors. The result is that scatter in $M_*$
``inflates'' the observed mass function at the high-mass end, a bias that becomes worse as the scatter increases (e.g.,
from additional \photoz{}-related error terms).  The goal in this work is to study evolution in the number density
distribution.  If the scatter term evolves with redshift, as would be expected because the S/N of observations degrades
with redshift, then the observed evolution may be biased by the changing importance of Eddington bias.

If the various $M_*$ error terms can be estimated, one solution is to perturb the final $M_*$ values until the scatter
is uniform across the sample.  For the \cat{}, we estimated the $M_*$ error for each galaxy resulting from the
uncertainty in the total magnitude estimate (which normalizes $M_*$).  If no \specz{} was available, we added in
quadrature to this value the expected $M_*$ error resulting from the assigned \photoz{} (according to the
redshift-dependent performance of the associated \photoz{} estimator as compared to \speczs{}).  Based on the maximum
errors obtained for galaxies in our sample, we set a target for the final uncertainty of all galaxies at $\sigma_{M_*}
= 0.115$ dex.  We used a Gaussian kernel with a width equal to the difference in quadrature between this target error
and the estimated error for each galaxy to describe the degree of perturbation required to make the final scatter
uniform for each $M_*$ estimate.  In other words, random draws from these kernels were added to each $M_*$ estimate to
obtain a set of perturbed $M_*$ estimates.  The scatter resulting from magnitude and redshift errors is uniform for
these perturbed values.  We did not account for the additional error term that arises from model-fitting uncertainties
in $M_*$ because these indicated no redshift dependence and are, themselves, uncertain.  The resulting mass functions
derived with the perturbed sample of $M_*$ values is presented in Section \ref{MF:scatter-normalized}.

A second solution to accounting for a varying Eddington bias is to assume an intrinsic shape for the $M_*$ or $L$
function and forward-model the data while accounting for the estimated uncertainties \citep[e.g.,][]{moutard16}.  As
described in \citet {leauthaud16} we consider the same sources of error on a per-galaxy basis as described above.  We
assume a double Schechter function \citep{baldry08} of the following form:

\begin{displaymath}
\phi(M_{*}) =  (\ln 10)\exp\left[-\frac{M_{*}}{M_{0}}\right] \times
\hspace{0.65\columnwidth}
\end{displaymath}
\vspace{-1ex}
\begin{equation}
\quad
\left\{ \phi_{1}10^{(\alpha_{1}+1)(\log M_{*} -\log M_{0})}
           + \phi_{2}10^{(\alpha_{2}+1)(\log M_{*} -\log M_{0})} \right\}
\end{equation}

%/Users/alexie/Work/Boss/Completeness/Pros/total_cmass_smf_schecter_params.txt

\noindent where $\alpha_{2}>\alpha_{1}$ and the second term dominates at the low mass end. We generate Monte Carlo
realizations of this function that sample various parameter ranges as described below.  A mock sample is drawn from
each realization and the individual scatter terms are added to $M_*$.  The mock samples are binned identically as the
data and compared to the observed number density distributions in an iterative approach that allows the input
parameters to be constrained.

\subsection{Sample Variance}\label{cosmic_variance}\label{sample_variance}

% Mock's described in Chapter_26 and do_mocks.pro

Large volume surveys significantly mitigate the impact of sample variance (often called ``cosmic variance'') which
arises from large-scale fluctuations in the spatial distribution of galaxies in the universe \citep[see][]{moster11}.
\citet{stringer09} show, for example, that galaxy surveys spanning more than $\sim$100 deg$^2$ are needed to overcome
sample variance on measurements of evolution in the mass function at $z < 1$.

An estimate of the sample variance in the \cat{} can be made using an abundance-matched mock catalog \citep[see][]
{leauthaud16}.  The volume of the mock, 1 Gpc$^3~h^{-3}$, can be divided into multiple sub-volumes corresponding to
0.1-width redshift slices of the 139.4 deg$^2$ \cat{}.  In each redshift bin, we study the mass function distribution
contributed from 4--5 mock sub-volumes with a similar volume to Stripe 82.  Additional observational errors as well
as redshift evolution are ignored.  In the $0.3 < z < 0.4$ bin (0.02 Gpc$^3~h^{-3}$), this experiment yields a
1-$\sigma$ error of 0.014 dex at $\log M_*/\msun \sim 11.0$, rising to 0.02 dex at $\log M_*/\msun \sim 11.6$.  For
$0.3 < z < 0.4$ (0.04 Gpc$^3~h^{-3}$), the value is 0.008 dex at $\log M_*/\msun \sim 11.0$ but remains at 0.02 dex for
$\log M_*/\msun \sim 11.6$.  The errors rise further towards 0.1 dex at $\log M_*/\msun \sim 12.0$ where Poisson errors
from the limited number of massive mock halos also contribute.

Our adopted sample variance and Poisson error estimates come from bootstrap resampling the derived number densities. We
divide the \cat{} into 214 roughly equal area regions and recompute number density functions after resampling with
replacement.  This technique yields consistent results as the mock catalog analysis with the benefit of allowing us to
map covariance matrices (see Appendix \ref{sec:covar}) that facilitate comparisons to theoretical predictions \citep
[see][]{benson14}.  Given the correlations in the large-scale clustering of dark matter
halos across halo mass, one expects strong covariance across $M_*$ and $L$ in galaxy number densities as inferred from
this analysis.

%% ----------------------------------------------------------------------------------------------------------------------------------
\bigskip
\section{Results}\label{results}

\subsection{Assumption-averaged estimate of the stellar mass function} % (fold)
\label{MF:best}

	% PLOT: mfn_best
	%   Made in mfnplots.pro in Chapter 33
	{\begin{figure*}[]
  		\epsscale{0.8}
	  	\plotone{\dir_plots/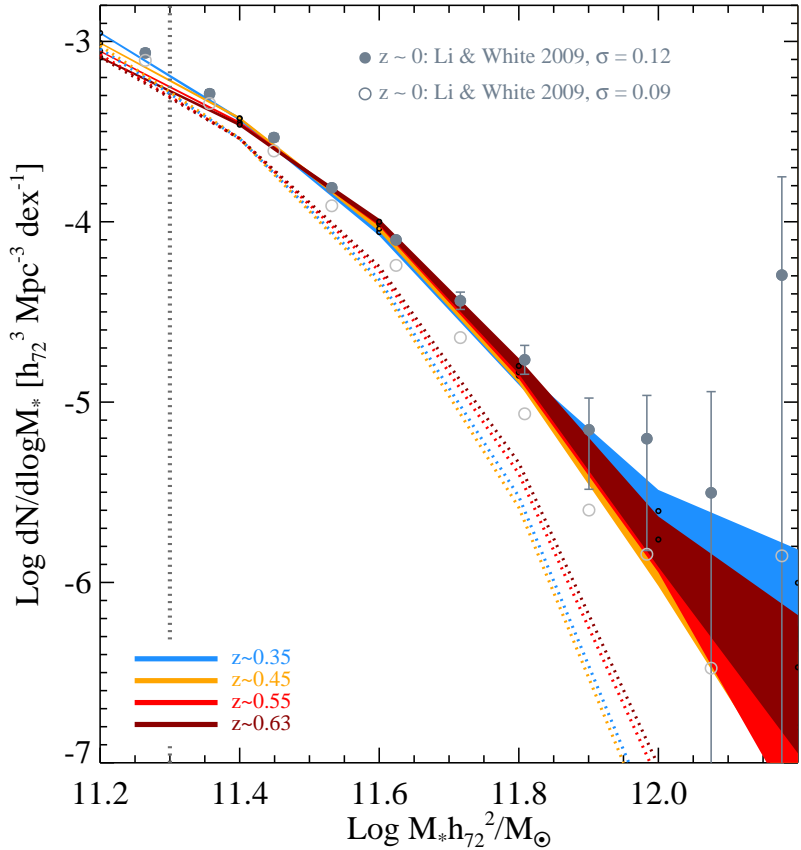}
		\caption{Assumption-averaged estimate $M_*$ function made by combining 4 separate $M_*$ estimators using different models and prior
		assumptions.  Shaded regions indicate Poisson errors only.  The estimated $M_*$ completeness is indicated by a
		vertical dotted line at $\log{M_*/\msun} = 11.3$.  Overplotted grey data points show the $z \approx 0$ SDSS MF and
		associated errors
		from \citet{li09}, after scaling their $M_*$ estimates to the \Mmgc{} (for galaxies in common) and convolving
		with two levels of scatter, as indicated.  Forward-modeling results, which aim to account for (and thereby
		remove) biases caused by measurement scatter, are shown with dotted lines.  These fits are subject to
		additional uncertainties in the assumed functional form and the modeling of various sources of scatter.}
		\label{fig:mfn:best}
	\end{figure*}}

We begin with estimates for the evolving galaxy $M_*$ functions derived from the \cat{} dataset after averaging
a set of four $M_*$ estimates made using different sets of priors and stellar population models.
In the sections that follow, we will examine how these functions change under different assumptions. Following
\citet{bundy15a}, we use the most accurate redshift available for each galaxy, $z_{\rm best}$, which is dominated by
\speczs{} for the majority of the sample.  Given subtle differences among $M_*$ estimates which we investigate below,
we define the ``assumption-averaged'' mass function from the average\footnote{In practice, the average
number densities are computed by binning a concatenated array of 4 different sets of $M_*$ estimates and dividing by
4 times the corresponding volume of each redshift slice.} of results from four different sets of 9-band $M_*$ estimates:
\Mmgc{} (original \cat{} estimates), \Mised{} (FSPS with bursts), \MisedBC{} (BC03 models, no bursts), and \MisedMa{}
(Maraston models, no bursts).  These four estimates encompass the range of $M_*$ values obtained by adopting currently
uncertain priors.  Without more information about how to set accurate priors or which models to favor, the
assumption-averaged result represents a compromise among differing approaches.

Figure \ref{fig:mfn:best} plots the ``as observed'' results with shaded regions corresponding to bootstrap errors
(i.e., both Poisson and sample variance errors are included).  No $M_*$ scatter normalization has been applied.  The
redshift bins are defined as $z = [0.3,0.4], [0.4,0.5], [0.5,0.6], [0.6,0.65]$ and we indicate the $M_*$ completeness
limit of $\log M_*/\msun = 11.3$ derived in \citet{bundy15a} with the vertical dotted line.

We have also forward-modeled the observed number densities to account for Poisson errors and scatter in $M_*$
uncertainties arising from SED fitting (fixed at 0.07 dex), \photoz{} uncertainty for galaxies without \speczs{}, and total flux errors, all of which are assumed to be Gaussian and are added in quadrature on a per-galaxy
basis.  A set of intrinsic fitted models are indicated as dotted lines with the same $z$-dependent colors.  Because the
modeling involves random draws from estimated error distributions, the intrinsic models can vary from run to run with a
scatter consistent with the error bars indicated on the raw mass functions in Figure \ref{fig:mfn:best}.

The modeling assumes the double-Schechter form described in Section \ref 
{method:scatter} and allows $\phi_{1}$, $\phi_{2}$ and $M_{0}$ to vary, while fixing $\alpha_{1}=-0.1$ and $\alpha_
{2}=-1.0$.  The choice of faint-end slopes and derived model parameters are degenerate and are not meant to convey
physical insight.  We have selected this model form because it accurately describes the data under our forward-modeling
analysis.  The results are given in Table \ref{tab:totalsmf_params}.  Tabulated data points are available from \href{http://www.massivegalaxies.com}{MassiveGalaxies.com}.

As in \comppaper{}, we can extend our characterization of galaxy stellar mass function to lower $M_*$ by including data
from other surveys.  For $M_* > 10^{10.4} \msun$, but below the completeness limit of the \cat
{}, our forward model fits include results from the PRIMUS mass functions \citep{moustakas13} observed at similar
redshifts.  While the PRIMUS data do not impact the derived mass functions at $M_* > 10^{11.3} \msun$, their inclusion
makes the intrinsic mass functions in Table \ref{tab:totalsmf_params} broadly representative of the galaxy
population with $M_* > 10^{10.4} \msun$ and $z < 0.6$. 

Within the statistically tight error bars from the \cat{} sample, we detect no redshift evolution over most of the mass
range probed.  At the lowest masses, there is a hint of positive growth (either in $M_*$ at fixed number density or in
number at fixed $M_*$), although this could likely reflect incompleteness at the faint-end, which would produce a
similar trend.  We will discuss the appropriate confidence level of our no-evolution result in Section \ref
{disc:noevol}.

The grey data points in Figure \ref{fig:mfn:best} represent the $z \approx 0$ mass function from  SDSS as derived by
\citet{li09}.  With smaller redshift surveys, comparisons to SDSS have been subject to systematic offsets in the
assumptions between $M_*$ estimates \citep[e.g.,][]{moustakas13}.  In the \cat{}, however, there are sufficient numbers
of galaxies that overlap with the \citet{li09} sample that we can characterize systematic offsets in $M_*$ and
statistically remove them.  The \citet{li09} $M_*$ estimates are taken from the Petrosian Kcorrect quantities which use
BC03 models and are provided in the NYU-VAGC \citep{blanton07}.  After adjusting the Hubble parameter to $h=72$, we
compare these \Mvagc {} values to \Mmgc{} for 3515 galaxies with $11.0 < \log M_*/\msun < 11.8$ and $0 < z < 0.2$.  We
fit a line to the mass difference ($\Delta \log M_* = \Mvagc - \Mmgc$) as a function of \Mvagc{}, referenced to $\log
\Mvagc/\msun = 11.3$, and adjust the \citet{li09} mass functions to account for the difference.  The fit's zeropoint
offset is 0.1 dex with a slope of -0.08.

Finally, we convolve the SDSS \citet{li09} mass function with additional scatter in $M_*$ to approximate the Eddington
bias in the \cat{} that results from larger photometric errors in both the total magnitudes and colors of the
higher-$z$ sample.  The convolution follows the approximation given in \citet{behroozi10}.  With typical total $K$-band
uncertainties of 0.05 mag, a reasonable estimate for the additional $M_*$ scatter in the \cat{} is $\sigma = 0.12$ dex.
Applying $\sigma = 0.12$ dex to the SDSS mass function results in the solid grey data points plotted in Figure \ref
{fig:mfn:best}.  The mass-adjusted \citet{li09} mass function with this additional scatter falls almost directly on the
\cat{} results, with a hint of lying on the more massive side of the \cat{} mass functions.   

However, our uncertainty in the correct amount of additional scatter to apply limits a precise comparison between the
\cat{} and SDSS $z \approx 0$ mass functions.  If we slightly reduce the applied scatter to $\sigma = 0.09$, still a
reasonable approximation to the true value, the resulting SDSS mass function falls significantly (0.1-0.2 dex) below
the \cat{} results.  We conclude that the \cat{} and SDSS $z \approx 0$ mass functions are in agreement, with no
detected differences at the 0.1 dex level.  This comparison includes a careful attempt to normalize the $M_*$
estimates, a process that should also remove biases from different estimators of total luminosity
\citep[e.g.,][]{bernardi13}.  However, a more precise treatment of $M_*$ scatter, let alone further assessments of
systematic biases in $M_*$ estimates (see below), is needed before these data sets can be used to measure growth in
$M_*$ with the needed sub-10\% level precision.

%- \textcolor{red}{ link to tables of the MF data}

\begin{deluxetable*}{@{}lccccc}
	\tablecaption{Intrinsic mass function shape parameters from forward modeling.}
	\tabletypesize{\footnotesize}
	\tablewidth{0pt}
	\tablecolumns{6}
	\tablehead{ \colhead{Redshift} & \colhead{$\log_{10}(\phi_{1}/$Mpc$^{-3}$dex$^{-1})$} & \colhead{$\log_{10}(\phi_
	{2}/$Mpc$^{-3} $dex$^{-1})$} & \colhead{$\log_{10}(M_0/M_{\odot})$} & \colhead{$\alpha_{1}$} & \colhead{$\alpha_{2}$} }

% Redshift & $\log_{10}(\phi_{1}/$Mpc$^{-3}$dex$^{-1})$ & $\log_{10}(\phi_{2}/$Mpc$^{-3} $dex$^{-1})$ & $\log_{10}(M_0/M_{\odot})$&$\alpha_{1}$  &$\alpha_{2}$ \\
	\startdata

$[0.30, 0.40]$         &   -5.92 $\pm$ 0.03   &   -2.50 $\pm$ 0.02   &   10.88 $\pm$ 0.01   &   -0.10  &   -1.00  \\
$[0.40, 0.50]$         &   -6.00 $\pm$ 0.03   &   -2.46 $\pm$ 0.01   &   10.87 $\pm$ 0.01   &   -0.10  &   -1.00  \\
$[0.50, 0.60]$         &   -5.63 $\pm$ 0.01   &   -2.60 $\pm$ 0.01   &   10.91 $\pm$ 0.01   &   -0.10  &   -1.00  \\
$[0.60, 0.65]$        &   -5.90 $\pm$ 0.02   &   -2.64 $\pm$ 0.01   &   10.93 $\pm$ 0.01   &   -0.10  &   -1.00  \\

	\enddata
	%\tablecomments{}
	\label{tab:totalsmf_params}

% $z_1=[0.15,0.43]$ &  -2.89     &   -2.84    &  10.89 &  -0.46& -1.58 \\
% $z_2=[0.43,0.54]$ &  -2.92     &   -2.92     &  10.92  & -0.46 &-1.58 \\
% $z_3=[0.54,0.63]$ &   -3.01    &   -2.94    &  10.98  & -0.46 & -1.58\\
% $z_4=[0.63,0.7]$ &   -3.05    &    -2.91    &   10.99  &  -0.46&-1.58\\
\end{deluxetable*}

\subsection{Scatter-normalized mass functions}
\label{MF:scatter-normalized}

	% PLOT: mfn_pert
	%   Made in mfnplots.pro in Chapter 33
	{\begin{figure}[h]
  		\epsscale{1.2}
	  	\plotone{\dir_plots/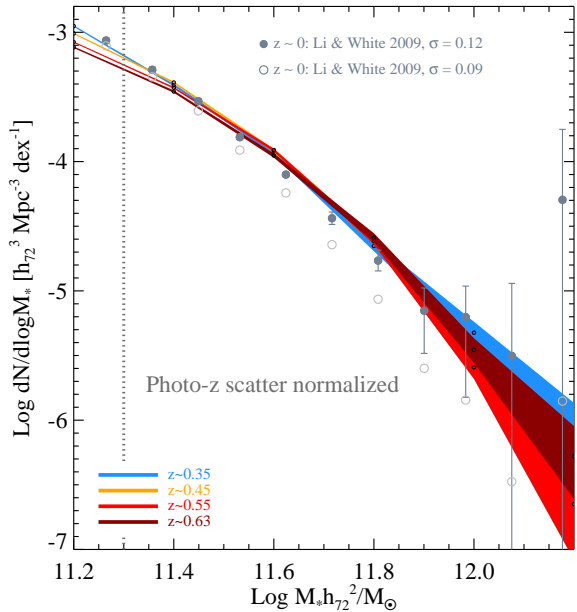}
		\caption{Mass functions as in Figure \ref{fig:mfn:best} but using $M_*$ estimates that have been perturbed to
		exhibit uniform \photoz-induced scatter across the redshift range probed.  The additional scatter causes an
		Eddington bias that inflates the derived number densities compared to Figure \ref{fig:mfn:best}, but this bias
		affects all redshift bins equally.  The scatter-normalized mass functions thus remain consistent with no
		evolution, confirming the results of the forward model fits in Figure \ref{fig:mfn:best}.  }
		\label{fig:mfn:pert}
	\end{figure}}

The assumption-averaged \cat{} mass functions, both in raw form and from forward model fitting, show no
evidence for redshift evolution.  While the forward model should account for the effect of scatter, we provide a
second test here using perturbed $M_*$ estimates.  Following the methodology in Section \ref{method:scatter}, we
perturb the $M_*$ values in order to normalize the scatter from \photozs{} and luminosity errors, aiming for a uniform
$\sigma_{M_*}$ uncertainty resulting from these two terms of 0.115 dex.  The mass functions using these perturbed $M_*$
values are shown in Figure \ref{fig:mfn:pert}.  As expected, the number densities are inflated with respect to Figure
\ref{fig:mfn:best}, but in a way that impacts all redshift bins equally.  The fraction of \photozs{} is relatively
small in the \cat{} and increases somewhat towards lower redshifts.  The combination of \photoz{} and luminosity error
in the $M_*$ uncertainties is thus roughly balanced as a function of redshift in the raw mass functions presented in
Figure \ref{fig:mfn:best}.

Confirming results from the previous section, no redshift evolution is apparent using the scatter-normalized $M_*$
values from the combined set of mass estimates.

\subsection{Dependence on priors} % (fold)
\label{MF:priors}

	% PLOT: mfn_priors
	%   Made in mfnplots.pro in Chapter 33
	{\begin{figure*}
  		\epsscale{1.2}
	  	\plotone{\dir_plots/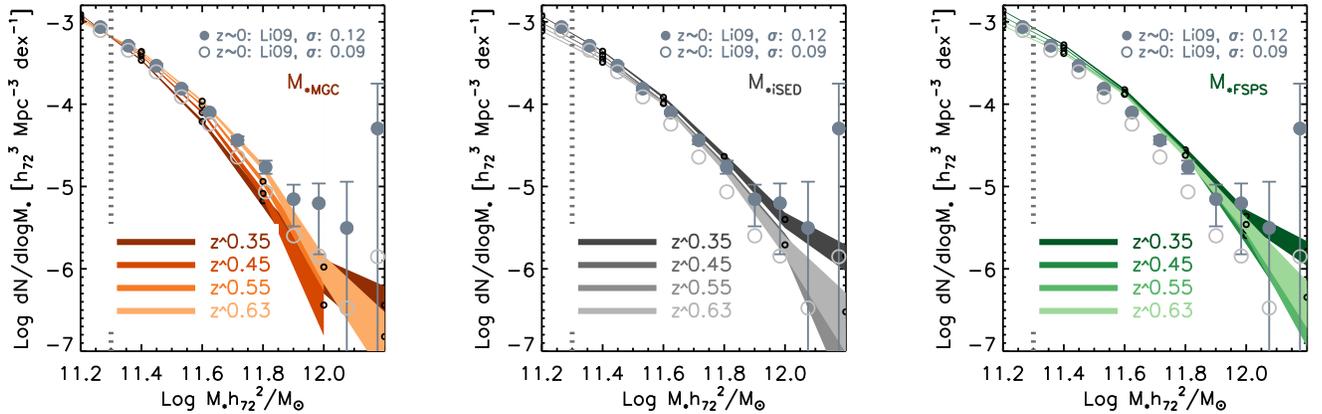}
		\caption{Mass functions obtained using three $M_*$ estimators with different star formation history prior assumptions.  The left panel
		corresponds to the fixed-grid priors of the \Mmgc{} estimates.  The resulting mass functions suggests a small
		($\sim$0.1 dex) decline in the average $M_*$ of massive galaxies over the redshift range plotted.  The trend is
		mildly reversed for \Mised{} masses (middle panel) based on FSPS and including bursts.  The right-most panel
		corresponds to \MisedFSPS {}, where the same models and global SFH priors have been assumed as in \Mised{} but
		no bursts are allowed.  Systematic }
		\label{fig:mfn:priors}
	\end{figure*}}

The mass function results from the previous sections average estimates from four different sets of $M_*$
measurements that include different star formation history priors and different stellar synthesis models.  These
assumption-averaged mass functions show no evidence for redshift evolution, but redshift differences do appear when
specific sets of $M_*$ estimates are used, underlying the importance of systematic errors in $M_*$ values when
measuring precise growth rates at these masses.  

We find that different priors in the star formation history lead to the largest discrepancies, both in terms of
absolute $M_*$ differences but more importantly in terms of the implied redshift evolution.  Figure
	\ref{fig:mfn:priors} shows raw mass functions based on three sets of $M_*$ estimates: \Mmgc{}, \Mised{}, and
	\MisedFSPS {}.  The \Mmgc{} mass functions (left panel) exhibit an apparent decrease of 0.1 dex in the $M_*$
	values of massive galaxies over the sampled redshift range.  Results with bursty star formation histories,
	(\Mised{}, middle panel) show a mild reversal of this trend, while the burst-free \MisedFSPS{} estimates (right
	panel) imply little to no evolution.  We show in the next section that the impact on evolutionary signals of
	different stellar synthesis models is modest, so while \Mmgc{} estimates are based on BC03 and the other estimates
	in Figure
	\ref{fig:mfn:priors} on FSPS, we ascribe most of the differences observed to star formation history priors.

The bursty \Mised{} mass functions (middle panel) not only suggest mild growth in $M_*$ with time---the opposite
conclusion of the \Mmgc{} results in the left-hand panel---but feature a more significantly elevated result at high
masses in the $z \approx 0.35$ bin compared to the \MisedFSPS{} number densities (right panel).  The \MisedFSPS{}
results are consistent with no evolution over the majority of the mass range probed.  The difference at higher masses
likely reflects the impact of priors that control the burst histories.  

While we leave a detailed investigation of the role of specific SFH priors and their optimization for this sample to
future work, we conclude from Figure \ref{fig:mfn:priors} that the resulting uncertainties introduce a systematic error
of 0.03 dex in the $M_*$ {\em growth} histories that we can determine from our combined assumption-averaged mass
function (absolute $M_*$ differences can be somewhat larger).  This level of systematic uncertainty resulting from
$M_*$ modeling is similar to that cited by \citet{moustakas13}.

\subsection{Impact of stellar synthesis models} % (fold)
\label{MF:models}

Figure \ref{fig:mfn:models} allows us to evaluate how three choices for the stellar population models underlying the
\isedfit{} $M_*$ estimates impact constraints on stellar mass growth.  In all cases, models without bursts are
compared.  The FSPS \MisedFSPS{} mass functions are repeated from Figure \ref{fig:mfn:priors} in the left-hand panel.
Mass functions based on BC03 masses,
\MisedBC{}, are shown in the middle panel, while the right-hand panel uses \MisedMa{}, based on models from
\citet{maraston05}.

From one panel to the next, \emph{absolute} differences in the mass estimates manifest in changes to the derived set of
mass functions.  But the implied differential redshift evolution within each panel is nearly identical, and again
consistent with no detectable growth with redshift.  At least among the set of stellar population synthesis models used
here, model differences are less important than star formation history priors in affecting conclusions about the
average growth rates in massive galaxy populations.

	% PLOT: mfn_models
	%   Made in mfnplots.pro in Chapter 33
	{\begin{figure*}
  		\epsscale{1.2}
	  	\plotone{\dir_plots/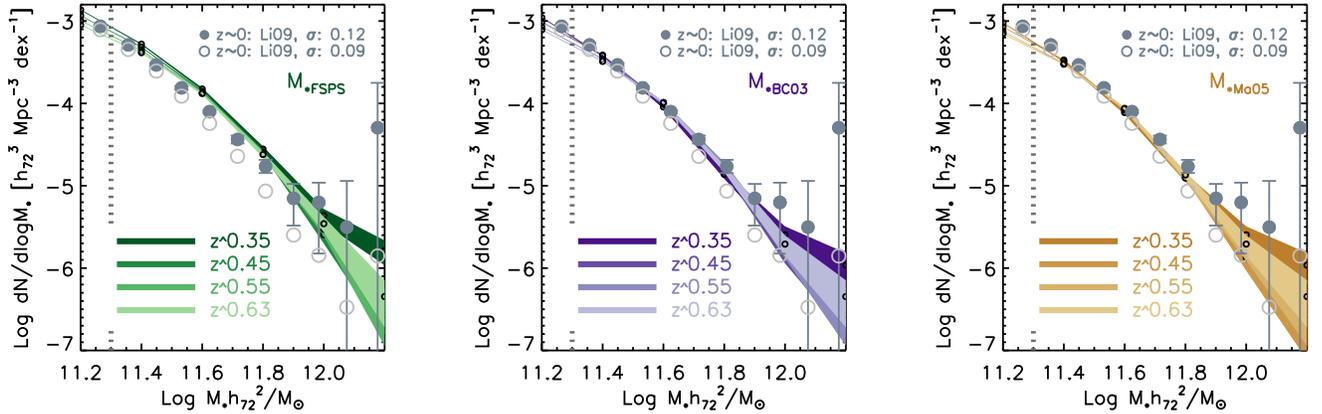}
		\caption{The impact of different stellar population synthesis models on the obtained mass functions.  All panels use
		\isedfit{} 
		$M_*$ estimates without bursts and the same SFH priors.  We compare \MisedFSPS{} (left panel), \MisedBC{}
		(middle panel), and \MisedMa{} (right panel) estimates.  The \emph{relative} differences as a function of
		redshift among the different stellar population synthesis models are sub-dominant compared to the impact of
		assuming different priors (Figure \ref{fig:mfn:priors}). }
		\label{fig:mfn:models}
	\end{figure*}}

\subsection{Dependence on star formation history} % (fold)
\label{MF:SFH}
	
	% PLOT: dist_SFH
	%   Made in call to plot_dist_SFH.pro in Chapter 33 under "color-dependent MFs"
	{\begin{figure}[h]
  		\epsscale{1.2}
	  	\plotone{\dir_plots/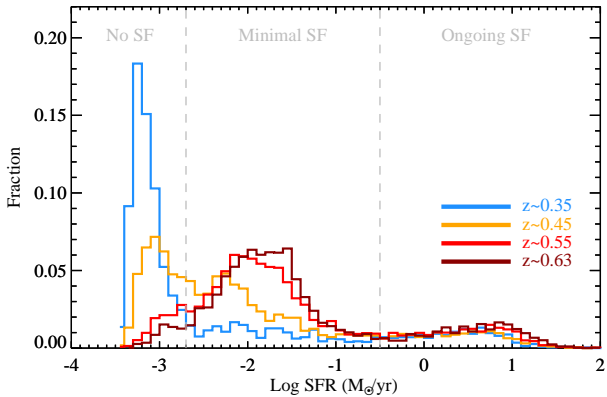}
		\caption{Distribution in star formation rates inferred from \isedfit{} in different redshift bins for galaxies
		with $\log M_*/\msun > 11.3$.  We classify galaxies into three groups as indicated by the vertical dashed
		lines. Systems with ongoing star formation, present at all redshifts, fall on the right-most side of the
		distribution.  A population with low, but possibly non-zero, star formation rates lies at the center.  This
		mid-SF population decreases with time.  On the far left, galaxies with the lowest derived SFR's, consistent
		with a complete lack of young stars, become increasingly abundant with time and dominate at the lowest
		redshifts.
		\label{fig:dist_SFH}}
	\end{figure}}

In this section, we partition the high-mass \cat{} galaxy population into different sub-samples based on the inferred
levels of recent star formation and investigate how the mass functions of these sub-samples evolve with time.  Our
information regarding each galaxy's star formation history (SFH) comes from fitting its SED to the 9-band \cat{}
photometry.  At the lowest redshifts we consider, $z=0.3$, the SDSS $u$-band samples the restframe near-UV, allowing us
to constrain the presence of young stars in a similar way as SDSS-I $z \approx 0.1$ studies employing UV data from
GALEX \citep[e.g.,][]{salim07}.  The near-IR bands help discriminate between reddening due to dust extinction versus
the red colors of aging stellar populations (see \catpaper{}).

Figure \ref{fig:dist_SFH} plots the redshift-dependent distribution of derived star-formation rates for \cat{} galaxies
with $\log M_*/\msun > 11.3$ using medians of the SFR posteriors reported by \isedfit{}.  The distribution of specific
star formation rates (sSFR) is qualitatively similar because of the narrow $M_*$ range of our sample, but is uniformly
low (these are passive galaxies).  We therefore focus on the un-normalized SFR given our interest on low-level,
residual star formation and the negligible impact such star formation has on $M_*$ growth for our sample.  With the
majority of SFR values below 1 \Msun/yr, their accuracy likely depends strongly on the SFH priors we have adopted,
which include a (poorly-constrained) prescription for bursts.  This is acceptable if our goal is to use these SFR
estimates as a proxy for examining broad differences in recent SFH across the high-mass population.  Other expressions
of these differences, such as the birth parameter, \birth{}, or stellar age, yield similar behavior.  With this in
mind, we divide the SFR distribution into three sub-samples.  We label galaxies with $\log {\rm SFR} < -2.7$ as having
``no star formation.'' Those with $-2.7 < \log {\rm SFR} < -0.5$ are interpreted as having experienced trace amounts of
recent star formation and labeled as ``minimally'' star-forming, while those with $\log {\rm SFR} > -0.5$ are
considered to have ongoing star formation.

The evolution of the $\log {\rm SFR}$ distribution suggests that our classification scheme may have physical meaning. 
At $z \approx 0.6$, Figure \ref{fig:dist_SFH} suggests that most high-mass galaxies are quiescent but had some minimal
recent star formation.  As time advances, this population declines and the majority of our sample falls into
the non-star-forming category.  It is interesting that this evolution suggests an exchange between two modes of
behavior as opposed to a smooth decrease in inferred SFR with time.  Meanwhile, a star-forming sub-sample remains
present and relatively consistent across the full redshift range.

	% PLOT: mfn_SFH
	%   Made in mfnplots.pro in Chapter 33
	{\begin{figure*}[h]
  		\epsscale{1.2}
	  	\plotone{\dir_plots/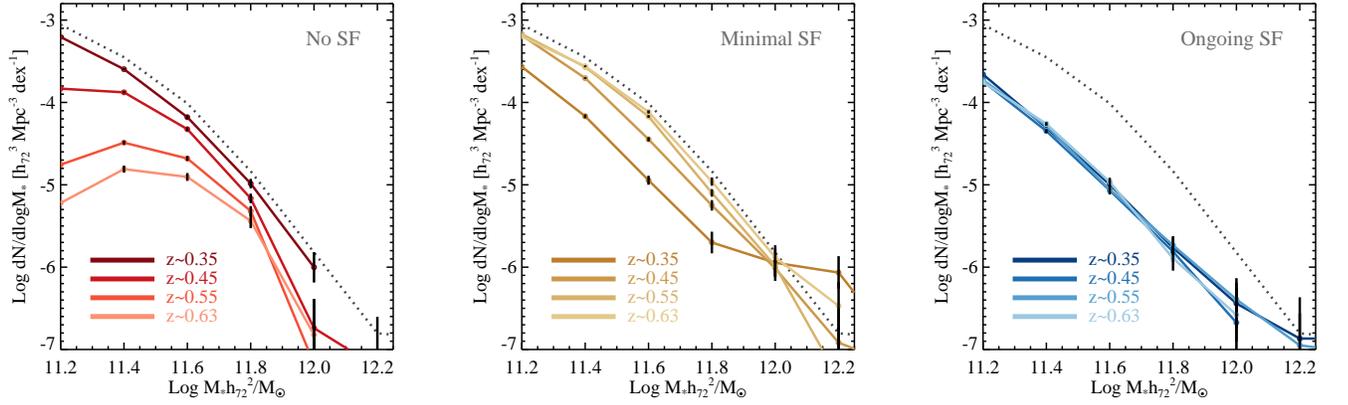}
		\caption{Evolving stellar mass functions of massive galaxies partitioned by the degree of recent star formation
		activity as derived from SED fitting.  Each panel corresponds to one of the three populations defined by cuts
		in the SFR distribution indicated on Figure \ref{fig:dist_SFH}).  The rising abundance of completely passive
		galaxies (left panel) as well as the declining numbers of minimally star-forming galaxies (middle panel) takes
		place at the lower end of the mass range studied in this work.  The highest masses (near 10$^ {12} \msun$) tend
		to be dominated by minimally star-forming galaxies at all redshifts.  Meanwhile, the mass function of galaxies
		with residual star formation hardly evolves and, interestingly, represents a greater fraction of the total
		population at the highest masses.  In all panels, the total mass function at $z \sim 0.55$ is plotted for
		comparison.  No corrections for scatter are applied to the plotted number densities. \label{fig:mfn:SFH}}
	\end{figure*}}

We can gain further insight by studying how the stellar mass functions of these SFH sub-samples evolve with time.  The
three panels in Figure \ref{fig:mfn:SFH} correspond to the ``no SF'', ``minimal SF'', and ``on-going SF'' populations.
Here we see that the evolutionary signal apparent in Figure \ref{fig:dist_SFH} is driven by galaxies at the
``lower-mass'' end of our sample, that is with $\log M_*/\msun \lesssim 11.8$.  The increase in the no-SF sample
coupled with the decline of the minimally star-forming populations at similar masses suggests an exchange, especially
given that the total mass function remains essentially fixed.  At the highest masses, $\log M_*/\msun \gtrsim 11.8$,
most galaxies remain in the minimally star-forming category at all redshifts.

The right-most panel of Figure \ref{fig:mfn:SFH} reveals the mass function of the star-forming population to be nearly
constant with time.  Its shape does not follow the total mass function but looks more like a power-law. Remarkably, we
see that the fraction of galaxies with ongoing star formation \emph{increases} at the highest masses, and while the
statistical uncertainties in our highest mass bin, $\log M_*/\msun = 12.2$, are too large to draw firm conclusions,
there is a hint that the majority of galaxies with such extreme $M_*$ estimates harbor a degree of ongoing star
formation at all redshifts.

% subsection dependence_on_star_formation_history (end)

%\subsection{Luminosity functions}

% Compare to Smith et al. 2009 (Smith, Anthony J.; Loveday, Jon; Cross, Nicholas
% J. G.) - mass and K-band luminosity functions at z~0 by combining SDSS-I and LAS

%% ----------------------------------------------------------------------------------------------------------------------------------
\section{Discussion}\label{discussion}

	\subsection{Confidence in Detecting No Evolution}\label{disc:noevol}

Even after accounting for $z$-dependent scatter, our assumption-averaged estimate of the high-$M_*$ mass function is consistent with
no evolution over $0.3 < z < 0.65$.  Here we summarize how different uncertainties affect this conclusion and limit the
degree of confidence associated with our claim of a lack of $M_*$ growth in the present analysis.

Poisson errors are essentially negligible, especially because any measure of $M_*$ growth would average the several
mass bins we sample at $\log M_*/\msun > 11.3$, while Poisson statistics are independent in each bin.  This is not the
case for remaining sample (``cosmic'') variance uncertainties which are highly covariant between mass bins (see Figure
\ref{fig:covar}).  Our mock \cat{} catalog suggests a 0.02 dex number density uncertainty for the mass function in the
smallest-volume, low-$z$ bin.  A number density deviation in one redshift bin at this level could be misinterpreted as
an 0.005 dex evolution in the average $M_*$.  It is unlikely that all four of our redshift bins would suffer
systematically increasing sample variance offsets, thereby conspiring to hide underlying $M_*$ growth.  Still, a
conservative estimate for the amount $M_*$ evolution that could be hidden would be a 2-$\sigma$ trend across redshift
amounting to 0.01 dex.

We have spent significant effort addressing concerns over the use of photometric redshifts, particularly their impact
on $M_*$ scatter (Section \ref{method:scatter}).  Regarding conclusions over global evolution, it is important to
emphasize that the spectroscopic completeness of these mass functions reaches 80\% above $\log M_*/\msun \approx 11.6$
(Chabrier IMF) even at the highest redshifts.  Systematic losses due to completeness are therefore an unlikely
contributor to our overall uncertainties.  The more general challenge of estimating the $M_*$ scatter could be
important, however.  In other words, it would be helpful to quantify the error on our error estimates.  In our effort
to make comparisons with the $z=0$ mass function, we noticed that the difference in assuming a total $M_*$ measurement
scatter of $\sigma = 0.12$ dex versus $\sigma = 0.09$ leads to changing mass function that could be misinterpreted as
implying 0.07 dex of $M_*$ evolution.  However, a $\sim$30\% systematic offset in our estimates of $\sigma$ versus
their true values seems unlikely over the well-detected high-mass galaxies in our redshift range.  A more reasonable
estimate for a potential systematic would be 0.02 dex.

In comparison to those above, the most significant systematic error we have studied so far are the potentially
$z$-dependent biases in $M_*$ estimates under different assumptions for star formation history (Section \ref{MF:SFH}).
Of the four different $M_*$ we combine in our assumption-averaged mass functions, the fiducial \Mmgc{} estimates (used alone)
would indicate a significantly measured decrease in $M_*$ over the redshift range.  The \Mised{} estimates employing
bursts would indicate a slight growth, while the BC03 \MisedBC{} and Maraston \MisedMa{} (neither with bursts) would
give no evolution.  Although a bursty SFH might be inconsistent with observed alpha-enhanced stellar populations
\citep[e.g.,][]{thomas05}, our current uncertainty in what priors to adopt leads us to combine these $M_*$ estimates
with equal weight and assess the resulting error on derived $M_*$ evolution to be 0.03 dex.

Combining these systematic error terms in quadrature yields 0.037 dex, suggesting that our results are consistent with
9\% or less evolution in the typical $M_*$ of high-mass galaxies over our redshift range.

There is one additional source of potential systematic error that will be addressed in future work and could dominant
over the 9\% estimate we quote above, namely a bias in our estimates of total luminosity.  We discuss this uncertainty
in more detail below.

\subsection{Biases from Luminosity Estimators}

	% PLOT: mfn_comp
	%   Made using mfnplots.pro in Chapter 33
	{\begin{figure*}[]
  		\epsscale{1.1}
	  	\plotone{\dir_plots/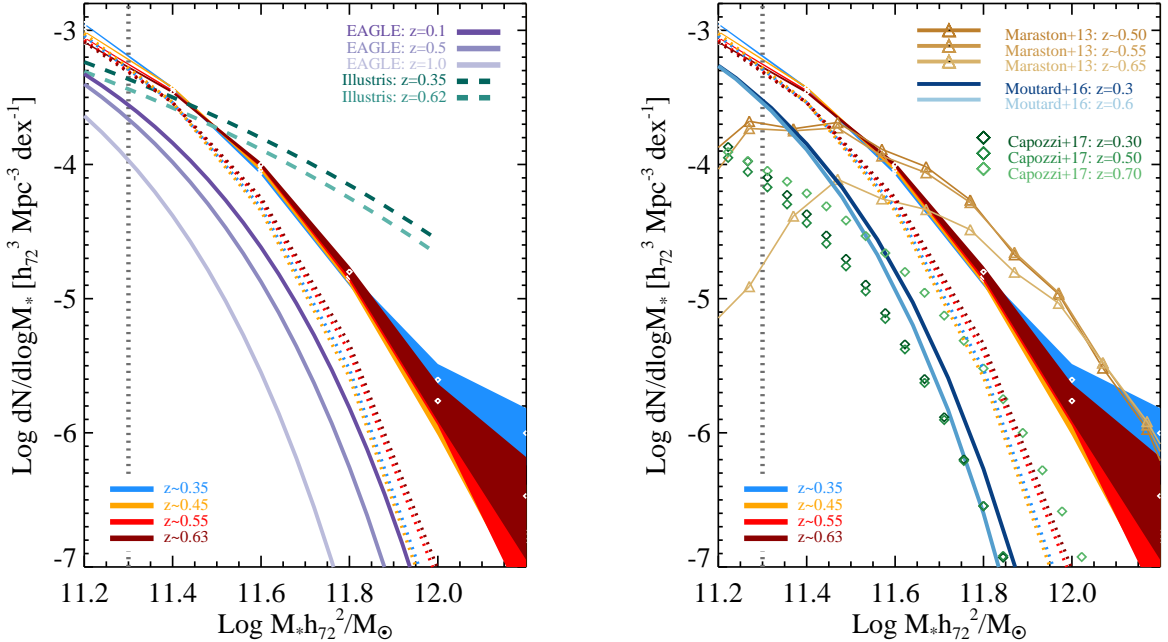}
		\caption{Comparison of mass function fits from cosmological hydrodynamic simulations (left panel) as well as previous observational
		results (right panel).  Both panels reproduce our assumption-averaged $M_*$ mass function results from Figure
		\ref{fig:mfn:best}, with the shaded regions indicating the raw number densities and associated error ranges and
		the thick, dotted lines representing the forward-model fitting results after accounting for measurement
		scatter.  On the left panel, the EAGLE mass functions from \citet{furlong15} should be compared to the
		forward-model results, while Illustris results from \citet{torrey17} should be compared to the raw number
		densities.  Simulations predict a $\sim$20\% growth in $M_*$ at fixed number density at these masses that is
		not observed.  On the right, we reproduce results \citep{maraston13} (based on BOSS) and \citet{capozzi17}
		(based on DES) that include measurement scatter.  The VIPERS-based \citet{moutard16} forward-models (blue
		curves) should be compared to our forward-model fits.  Global offsets in $M_*$ values from different estimators are expected; the sense and strength of claims of internal redshift evolution are of particular interest.
		\label{fig:mfn:comp}}
	\end{figure*}}

Stellar mass estimates ultimately rely on measures of the total luminosity of galaxies.  Even with $z \approx 0$ SDSS
samples, choices in how surface brightness profiles are fit can have dramatic implications for derived $M_*$ estimates
and resulting stellar mass functions \citep{bernardi13}.  At the highest masses, discrepancies in $M_*$ estimates can
reach two orders of magnitude, depending on profile fitting assumptions \citep{bernardi17,huang17}.  

Detailed work on nearby galaxies has emphasized the multi-component nature of galaxy light profiles---spheroidal
galaxies often exhibit an outer component that, while low in surface-brightness, can contribute significantly to total
$M_*$ \citep{huang13}. There is evidence that the outer components of the most massive central galaxies grow with time
even since $z \sim 0.6$ \citep {vulcani14} and that their rising importance accounts for a degree of claimed size
evolution \citep[e.g.,][]{van-der-wel14}.

Indeed, studies of the evolving mass-size relation put a premium on deep photometry, often from the Hubble Space
Telescope, and pay close attention to biases in 2D profile fitting.  Unfortunately, the photometric data sets that
underlie galaxy redshift surveys on which number density studies are often based (including this one) are much
shallower.  Photometry requirements are typically just deep enough to detect galaxies in the sample, not to
characterize their low surface-brightness outskirts.  \citet{tal11} use stacking analyses, for example, to show that
SDSS imaging misses 20\% of the total light of luminous red galaxy (LRG) samples.  Shallow imaging depths also motivate
the use of rather simple total luminosity estimators, such as the Kron and Hall estimators that underlie our $M_*$
estimates in the
\cat{}.

We therefore consider a major limitation of this work our inability to quantify the stellar content of the outer
components of massive galaxies.  Future work exploiting deeper data sets like the Hyper Suprime-Cam Survey may reveal
significant growth in these components, which remain below the detection level of the \cat{} even at the lowest
redshifts probed.  It is possible that their presence could have a profound affect on conclusions regarding evolution
in the total mass function.

	\subsection{Comparisons to Other Results}

Figure \ref{fig:mfn:comp} presents a comparison of the \cat{} mass functions to both theoretical results (left panel)
and recent observational work spanning large volumes (right panel).  In both cases, we reproduce from Figure
\ref{fig:mfn:best} the raw number counts from the assumption-averaged \cat{} mass function with associated error bars indicated by
shaded regions as well as fits from forward-modeling the raw results (thick, dotted lines).  The forward models account
for our estimates of various sources of measurement error.

We compare to theoretical results from recent cosmological hydrodynamic simulations (left panel). Stellar mass
functions for the EAGLE simulation are taken from fits provided at specific redshifts by \citet{furlong15}. For a
comparison to the Illustris Simulation results we use the mass function fitting formulae provided in \citet{torrey17}
and evaluate the relation at the midpoint of the lowest and highest redshift bins in the \cat{} sample.  Both
simulations predict a $\sim$20--30\% growth in $M_*$ at fixed number density at these masses that is not detected in
our data.  For a direct comparison to Illustris, the raw mass functions may be appropriate as the Illustris output was
tuned to reproduce the evolving galaxy stellar mass function, as observed at lower $M_*$ \citep{torrey14}.  These
observational results likely included the effects of measurement scatter, which would be expected at $z \gtrsim 0.3$ to
be similar to the uncertainties estimated here.  We see, however, that the Illustris number densities, while in broad
agreement with the \cat{} at $\log M_*/\msun \approx 11.5$, trace a shallower mass function than what we observe and
land an order of magnitude too high at 10$^{12}$ \Msun{}, although they are in closer agreement with \citet{maraston13}
(see below).  At this $M_*$, \citet{torrey17} warn that Illustris becomes incomplete.

The EAGLE simulation output was tuned to the SDSS-based $z \approx 0.1$ mass function alone which, being at low
redshift, suffers from less measurement error.  A direct comparison with EAGLE is therefore more appropriately made
with our forward model fits in which we have attempted to remove the effects of scatter.  Here the agreement with our
observations in both shape and normalization is better.  If applied to the EAGLE results, a constant $M_*$ offset of
+0.05 dex, well within expectations for mass estimator differences, would bring the low-$z$ mass functions into
agreement, and \citet{furlong15} speculate that galaxies in their simulation may be over-quenched.  Our forward-model
results, however, are inconsistent with the smooth redshift evolution predicted by EAGLE (see Section
\ref{disc:noevol}).

The right-hand panel of Figure \ref{fig:mfn:comp} compares our results to other observational efforts.  The raw number
densities derived from the \citet{maraston13} analysis of the BOSS sample are overplotted with open symbols connected
by gold lines.  Corrections to $h$ have been applied, but \citet{maraston13} do not account for scatter and so should
be compared to the raw number counts from the \cat{}.  In \catpaper{}, we show that \Mmgc{} is systematically larger
than the \citet{maraston13} $M_*$ estimates, an offset that increases with $M_*$ to 0.1 dex at $\log M_*/\msun \sim
11.8$.  The higher \citet{maraston13} number densities at fixed $M_*$ in Figure \ref{fig:mfn:comp} may owe instead to
Eddington bias from larger $M_*$ uncerainties \citep{bundy15a}.  Note the effect of CMASS sample incompleteness in the
highest redshift bin from \citet{maraston13}.  

The DES \photoz-only Schechter fits from the mass functions in \citet{capozzi17} are overplotted as green open symbols.
These are fits to raw number counts (no scatter correction) and should be compared to our raw number densities (shaded
curves). On top of a global $M_*$ offset\footnote{The $M_*$ estimates in \citet{capozzi17} use a Salpeter IMF, which
introduces a +0.25 dex offset compared to the Chabrier-based values in this paper.  However, the adopted SF priors in
\citet{capozzi17} were shown by \citet{maraston13} to cause a -0.25 dex offset.  Since this cancels the offset from the
Salpeter IMF compared to our estimates, we plot the \citet{capozzi17} results without any corrections applied.}, the
\citet{capozzi17} results favor a decreasing mass function with time that would be consistent with a decrease in the
typical $M_*$ of massive galaxies.

Finally we overplot the forward-model results of \citet{moutard16} (solid blue lines) which are based in part on VIPERS
data and should be compared to our forward-model fitting approach (with scatter removed).  Acknowledging a global $M_*$
offset, the evolutionary signal claimed by \citet{moutard16} appears to have a similar amplitude as the range in
forward-modeled mass functions that we derive.  Given the uncertainties in our data, we do not interpret this range to
be physically meaningful.  Separate modeling runs with different random draws of our error distributions yield
different relative orientations of our redshift-dependent mass functions.  With 22 deg$^2$ compared to our 139 deg$^2$,
the \citet{moutard16} data set may have similar (or greater) uncertainties.  The apparent evolution in their mass
function fits might therefore arise from differing priors on the mass function shape parameters.

	\subsection{Dependence on Star Formation History}

The SFR distributions presented in Figure \ref{fig:dist_SFH} suggest that massive galaxies can be classified according
to the degree of low-level star formation that is present.  Figure \ref{fig:mfn:SFH} shows that the population with
some residual star formation stays constant with time, while the abundance of galaxies with minimal star formation 
decreases, apparently resulting in a build-up of systems with no star formation at all.  These results are based on the
optical-near-IR fitting we have performed with \isedfit{} and therefore reflect features in broadband SEDs.  They are
also subject to the adopted priors which, for example, limit derived SFRs to be greater than $\sim$10$^{-3}$ \Msun
{}/yr likely resulting in the apparent peak at this value in Figure \ref{fig:dist_SFH}.

Is the apparent decline in the abundance of high-mass galaxies with minimal star formation real?  If so, it may be a
sign-post of more recent quenching, past merging episodes with smaller, gas-rich galaxies or low levels of residual gas
cooling and star formation that become increasingly rare towards the present day.  Alternatively, could the global
shift towards near-zero SFRs simply reflect passive evolution of exponentially-declining SFHs?

Constraints on star formation histories from detailed analysis of massive galaxy spectra present a complementary view
\citep[e.g.,][]{thomas05,tojeiro07,thomas10}.  For $\log M_*/\msun < 11.5$, \citet{choi14} stack $z \sim 0.5$ spectra
to argue that more massive quiescent galaxies have older SSP-equivalent ages at all times (for $z \lesssim 1$).
However, while this age-mass trend generally evolves towards older ages with time, the lower-mass galaxy populations
age less rapidly.  This suggests more complex SFHs, possibly resulting from recent red-sequence arrivals which may also
contribute to the ``minimal'' SFR population we identify here.

\citet{choi14} present exponential SFHs that are meant to globally capture the mass and age trends of their stacked
samples.  The data are broadly consistent with a short burst ($\tau = 0.1$ Gyr) of star formation at $z \sim 1.2$ as
well as with longer declining histories ($\tau \sim 2$ Gyr) initiated at $z = 3$.  Neither of these global models
explain the SFR distributions we see in Figure \ref{fig:dist_SFH}.  Short bursts at $z \sim 1$ have completely
extinguished by $z \sim 0.5$, and even if our absolute measure of SFR could be made consistent, the longer SFH models
predict 0.2-0.3 dex of gradual decline in SFR per 0.1-wide redshift bin.  Our estimates suggest a much more dramatic
cessation.  While the exponential models may provide a useful description for the majority of stars in massive
galaxies, we conclude that residual low-level star formation may still be present in ways that shed light on recent
assembly history.

We turn now to the more rare phenomenon of very massive galaxies exhibiting significant levels of star formation, with
SFR $\gtrsim 1$ \Msun/yr.  One concern is that our SED-based SFR estimates are biased by a ``UV upturn'' which is
likely a signature of stellar evolution, not a sign of recent star formation.  Figure 13 in \catpaper{} shows how a
related measure of recent star formation, the \birth{} parameter, varies across the optical-near-IR color space of our
sample.  This plot demonstrates that the majority of our modestly star-forming galaxies have red optical colors.  The
near-IR photometry is what allows us to distinguish them as (mildly dust-obscured) star-formers, not a detection of
enhanced UV flux. We also reported that the visual morphologies of these galaxies are predominantly disk-like or
disturbed.

This star-forming population remains remarkably constant across our redshift range.  From $10^{11}$ to $10^{12}$ \Msun,
roughly 10\% of massive galaxies are in throes of a noticeable star-forming episode.  While unlikely to build
significant additional stellar mass, these episodes may again be signs of an active (minor) merging history which in
some cases may significantly revive quiescent galaxies \citep[e.g.,][]{kannappan09}.  At $10^{12}$ \Msun{} and above,
galaxies with significant star formation appear to be far more common.  Assuming that most of these systems are
Brightest Cluster Galaxies (BCGs), these results are consistent with BCG sample studied by \citet{mcdonald16}.  They
find that 34\% of BCGs at $0.25 < z < 1.25$ have SFR$ > 10$ \Msun/yr.  At $z < 0.6$, \citet{mcdonald16} use entropy
profiles in the hot intracluster medium (ICM) to argue that cooling in relaxed cool-core clusters provides the dominant
source of SFR fuel. The rising fraction of star-formers we see in our sample towards higher $M_*$ may signal the
increasing role of ICM cooling in triggering high-$M_*$ star formation.  Unfortunately, statistical uncertainties limit
our ability to study evolution in the abundance of $M_* > 10^{12} \msun$ star-formers, and we do not probe beyond $z
\sim 0.6$ where \citet{mcdonald16} argue ICM cooling no longer correlates with BCG star formation.

%% ----------------------------------------------------------------------------------------------------------------------------------
\section{Summary and Conclusions}\label{summary}

We have exploited optical through near-IR matched photometry in the \cat{} to measure the galaxy stellar mass function
in four redshift bins from $z = 0.30$ to $z = 0.65$.  While our $M_*$ completeness of $\log M_*/\msun > 11.3$ is
relatively shallow, our sample spans a large area of $139$ deg$^2$, delivering exquisite statitical precision on
possible evolution at the highest masses.

We pay special attention to sources of random and systematic error and investigate their effects on our derived mass
functions through both forward modeling and perturbations to our measurements that result in samples with uniform
measurement uncertainties.  The two techniques yield consistent results.  These techniques also address concerns from
the use of photometric redshifts, although our sample has a high degree (80\%) of spectroscopic redshift completeness,
even at the highest redshifts we probe.

Our key result is shown in Figure \ref{fig:mfn:best}.  After combining $M_*$ estimates that adopt a range of currently
uncertain prior assumptions, we find no evolution in the typical $M_*$ at fixed number density for massive galaxies in
our redshift range.  Recent simulations predict growth in $M_*$ of 20-30\%.  Taking account of errors studied in this
work, we can rule out evolution in $M_*$ of 9\% or more.  Among those considered here, the largest contribution to this
uncertainty are biases in $M_*$ estimates arising from different SFH priors.  However, we speculate that missing light
from our adopted total luminosity estimators is of far greater importance and, when accounted for in future work, could
strongly impact our conclusions.

Finally, we divide our sample based on the degree of resiudal, low-level star formation as determined from our SED
fitting.  We find a minimally star-forming population that appears to become completely passive over our redshift
range. There is an additional less abundant population with notable, but still low SFR (about 1 \Msun/yr) whose mass
function hardly evolves.  Interestingly, this population becomes more common at the highest masses and may be
associated with brightest cluster galaxies in cool-core clusters.

%% ----------------------------------------------------------------------------------------------------------------------------------
\section{Acknowledgments}

	This work was supported by World Premier International Research Center
	Initiative (WPI Initiative), MEXT, Japan.  This work was supported by a
	Kakenhi Gant-in-Aid for Scientific Research 24740119 from Japan society
	for the Promotion of Science.  We thank E.~Rykoff and E.~Rozo for a
	generous contribution of {\it RedMapper} photometric redshift estimates.
	This publication has made use of code written by James R. A. Davenport.

\bibliographystyle{apj}
\bibliography{/Users/kbundy/Documents/bibliographies/references.bib}

\appendix
\section{Covariance Matrices}
\label{sec:covar}

\begin{figure}
    % \epsscale{1.2}
    % \plottwo{\dir_plots/mfn_covar_0.eps}{\dir_plots/mfn_covar_1.eps}
    \centering
    \includegraphics[height=5cm]{\dir_plots/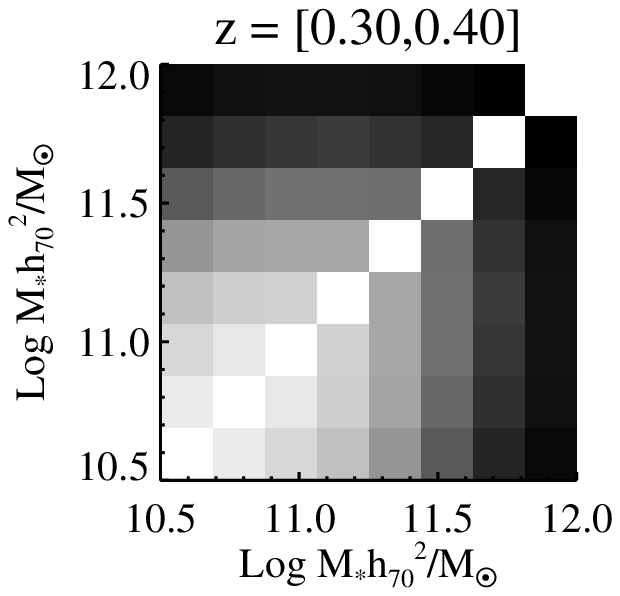}
    \includegraphics[height=5cm]{\dir_plots/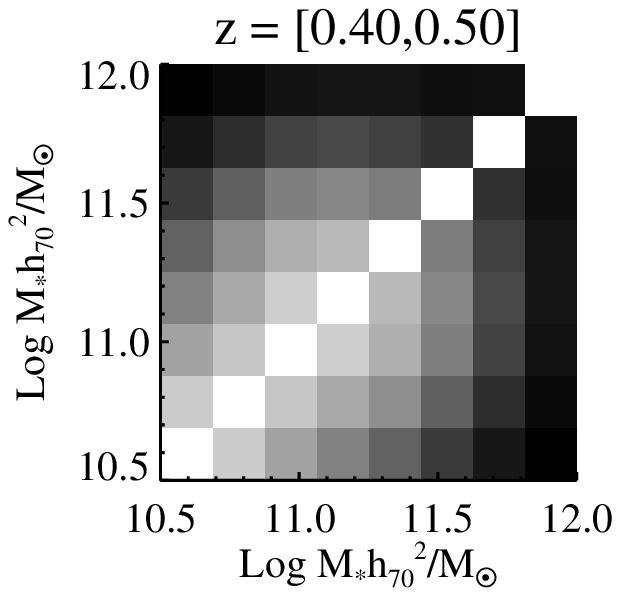} \\
    \includegraphics[height=5cm]{\dir_plots/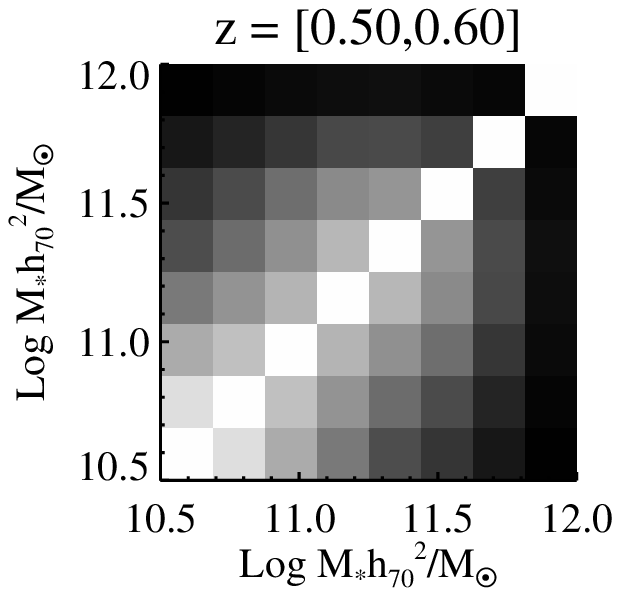}
    \includegraphics[height=5cm]{\dir_plots/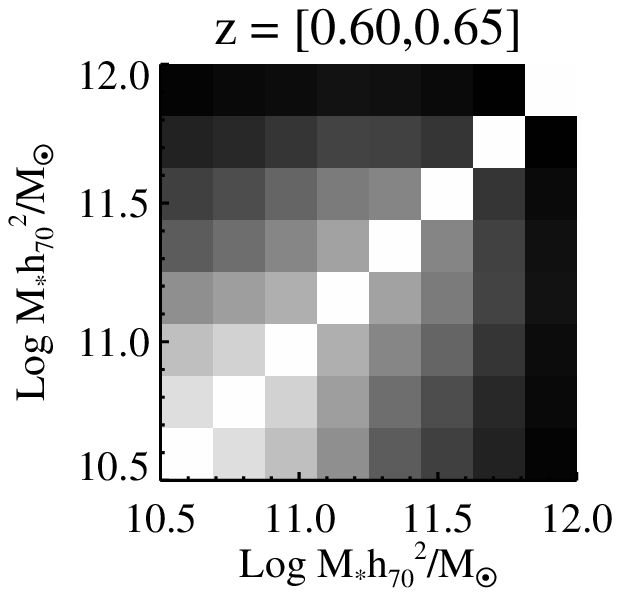}
    \caption{Correlation matrices from the normalized covariance of the \cat{} mass functions as determined from
    gridding the survey footprint into 214 subregions and resampling with replacement.\label{fig:covar}}
\end{figure}

The correlation matrices from the bootstrap resampling are plotted in Figure \ref{fig:covar} and made available at {\href{http://www.massivegalaxies.com}{MassiveGalaxies.com}}.

\end{document}